\documentclass{pasj00}
\twocolumn

\usepackage{natbib}

\begin{document}
\SetRunningHead{Nagamine et al.}{LAEs and LBGs at $z=3-6$}
\Received{2010/07/07}
\Accepted{2010/09/28}

\title{Lyman-$\alpha$ Emitters and Lyman-break Galaxies at $z=3-6$ in Cosmological SPH Simulations}

\author{Kentaro \textsc{Nagamine} %
  \thanks{Visiting Researcher, Institute for the Physics and Mathematics of the Universe, University of Tokyo, 5-1-5 Kashiwanoha, Kashiwa, 277-8568, Japan}}
\affil{University of Nevada Las Vegas, Department of Physics \& Astronomy, 4505 Maryland Pkwy, Box 454002, Las Vegas, NV 89154-4002 U.S.A.} 
\email{kn@physics.unlv.edu}

\author{Masami \textsc{Ouchi}
\thanks{Current address: Institute of Cosmic Ray Research, University of Tokyo, 5-1-5 Kashiwa-no-ha, Kashiwa, Chiba, 277-8582, Japan}}
\affil{Observatories of the Carnegie Institution of Washington, 813 Santa Barbara Street, Pasadena, California, 91101 U.S.A.}

\author{Volker \textsc{Springel}
\thanks{Current address: Heidelberg Institute for Theoretical Studies, gGmbH, Schloss-Wolfsbrunnenweg 35, D-69118 Heidelberg, Germany}}
\affil{Max-Planck-Institut f\"{u}r Astrophysik, Karl-Schwarzschild-Stra\ss{}e 1, 85740 Garching bei M\"{u}nchen, Germany}
\and
\author{Lars {\sc Hernquist}}
\affil{Harvard University, 60 Garden Street, Cambridge, MA 02138, U.S.A.}


\maketitle

\newcommand{\Mstar}{M_{\star}}
\newcommand{\Mbh}{M_{\rm BH}}
\newcommand{\Mhalo}{M_{\rm halo}}
\newcommand{\Muv}{M_{\rm UV}}
\newcommand{\Lam}{\Lambda}
\newcommand{\lam}{\lambda}
\newcommand{\Del}{\Delta}
\newcommand{\del}{\delta}
\newcommand{\mpc}{\rm Mpc}
\newcommand{\kpc}{\rm kpc}
\newcommand{\yr}{\rm yr}
\newcommand{\km}{\rm km}
\newcommand{\s}{\rm s}
\newcommand{\kms}{\rm km\,s^{-1}}
\newcommand{\erg}{\rm erg}
\newcommand{\Msun}{M_{\odot}}
\newcommand{\Lsun}{L_{\odot}}
\newcommand{\Zsun}{Z_{\odot}}
\newcommand{\hinv}{h^{-1}}
\newcommand{\himpc}{\hinv{\rm\,Mpc}}
\newcommand{\hikpc}{\hinv{\rm\,kpc}}
\newcommand{\himsun}{\,\hinv{\Msun}}

\newcommand{\Om}{\Omega_{\rm m}}
\newcommand{\Ol}{\Omega_{\Lam}}
\newcommand{\Ob}{\Omega_{\rm b}}
\newcommand{\OHI}{\Omega_{\rm HI}}
\newcommand{\HI}{H{\sc i}\,\,}
\newcommand{\NHI}{{N_{\rm HI}}}
\newcommand{\Mtot}{{M_{\rm tot}}}
\newcommand{\Lbox}{{L_{\rm box}}}
\newcommand{\Csto}{{C_{\rm stoc}}}
\newcommand{\Lya}{Ly$\alpha$}
\newcommand{\La}{L_{\rm Ly\alpha}}
\newcommand{\fLya}{f_{\rm Ly\alpha}}
\newcommand{\fesc}{f_{\rm esc}}
\newcommand{\fion}{f_{\rm esc}^{\rm ion}}
\newcommand{\figm}{f_{\rm IGM}}
\newcommand{\highz}{high-$z$}
\newcommand{\fdust}{f_{\rm dust}}
\newcommand{\SFR}{{\rm SFR}}
\newcommand{\lgZ}{\log (Z/Z_\odot)}

\begin{abstract}
  We study the properties of Lyman-$\alpha$ emitters (LAEs) and Lyman-break
  galaxies (LBGs) at $z=3-6$ using cosmological SPH simulations.  We
  investigate two simple scenarios for explaining the observed \Lya\ and
  rest-frame UV luminosity functions (LFs) of LAEs: (i) the ``{\em escape
    fraction}'' scenario, in which the {\it effective} escape fraction
  (including the IGM attenuation) of \Lya\ photons is $\fLya \approx 0.1$
  (0.15) at $z=3$ (6), and (ii) the ``{\em stochastic}'' scenario, in which the
  fraction of LAEs that are turned on at $z=3$ (6) is $\Csto\approx 0.07$
  (0.2) after correcting for the IGM attenuation.  Our comparisons with a
  number of different observations suggest that the stochastic scenario is
  preferred over the escape fraction scenario.  We find that the
  mean values of stellar mass, metallicity and black hole mass hosted by LAEs
  are all smaller in the stochastic scenario than in the escape fraction
  scenario.
  In our simulations, the galaxy stellar mass function evolves rapidly, as
  expected in hierarchical structure formation. However, its evolution is
  largely compensated by a beginning decline in the specific star formation
  rate, resulting in little evolution of the rest-frame UV LF from $z=6$ to 3.
  The rest-frame UV LF of both LAEs and LBGs at $z=3$ \& 6 can be described 
  well by the stochastic scenario provided the extinction is moderate, 
  $E(B-V)\approx 0.15$, for both populations, although our simulation might 
  be overpredicting the number of bright LBGs at $z=6$.  
  We also discuss the correlation function and bias of LAEs. 
  The \Lya\ LFs at $z=6$ in a field-of-view of $0.2\,\deg^2$ show a 
  significantly larger scatter owing to cosmic variance relative to that 
  in a $1\,\deg^2$ field, and the scatter seen in the current observational 
  estimates of the \Lya\ LF can be accounted for by cosmic variance.
 \end{abstract}

\section{Introduction}

Hydrogen is ubiquitous in our universe, and its \Lya\ emission line is 
now commonly observed in the spectra of \highz\ galaxies
\citep[e.g.,][]{Shapley03, Ouchi05b, Hu06}.  Recent detections of LAEs in
large numbers at $z=3-6$ using narrow-band filters have opened up a new 
window for studying \highz\ galaxies after the long effort since the 
original proposal \citep{Partridge67} of using \Lya\ emission for the 
search of \highz\ galaxies.

In particular, wide-field surveys have been extremely successful in 
searching for LAEs, and the sizes of observed samples are becoming 
comparable to that of LBGs with several hundred sources 
\citep[e.g.,][]{Ouchi03a, Hu04, Dawson04, Malhotra04, Ouchi05a, 
Taniguchi05b, Venemans05, Shimasaku06, Kashikawa06b, Gronwall07, 
Murayama07, Ouchi10}.  These large samples allow us to construct the luminosity 
function (LF) of LAEs with reasonable accuracy and to perform statistical 
analyses and comparisons with other populations such as LBGs, DRGs, 
EROs, BzKs, etc.

At the same time, deep observations of smaller samples at infrared (IR)
wavelengths are beginning to constrain the physical properties of LAEs 
such as stellar mass and star formation rate (SFR).  For example, 
\citet{Gawiser06} stacked the SEDs of 18 LAEs at $z\simeq 3.1$, and 
estimated an average stellar mass of 
$\Mstar \simeq 5\times 10^8\,\Msun$ and a star formation 
rate of $\SFR \simeq 6\,\Msun\,\yr^{-1}$ by fitting the
SED using a population synthesis model.  \citet{Gawiser07} updated the 
result using 162 LAEs from the MUSYC survey and reported $\Mstar \simeq
1.0^{+0.6}_{-0.4}\times 10^9\,\Msun$ and 
$\SFR \simeq 2\pm 1\,\Msun\,\yr^{-1}$ for typical LAEs at $z=3.1$.  
\citet{Lai08} analyzed the same sample of $z=3.1$ LAEs supplemented by 
the Spitzer IRAC observations, and found $\Mstar
= 3^{+4}_{-2}\times 10^8\,\Msun$, an age of $\sim 200$\,Myr, and an 
average SFR of $2\,\Msun\,\yr^{-1}$ for the IRAC(3.6$\mu$m)-undetected 
sample.  The IRAC-detected sample is more massive with 
$\Mstar = 9\pm 3 \times 10^9\,\Msun$, ${\rm age}=1.6\pm 0.4$\,Gyr, 
and ${\rm SFR}\sim 6\,\Msun\,\yr^{-1}$.

At higher redshifts, \citet{Pentericci07} studied $z\sim 4$ LBGs with 
and without \Lya\ emission, and concluded that the LBGs with \Lya\ 
emission are on average much younger and less massive than the LBGs 
without \Lya\ emission. They estimated $\Mstar = (5\pm 1) \times 10^9 
\Msun$ and an age of $200\pm 50$\,Myr for the LBGs with \Lya\ emission.
\citet{Pirzkal07} studied the SEDs of 9 LAEs in the Hubble Ultra Deep 
Field at $4.0<z<5.7$, and estimated $\Mstar=10^6 - 10^8 \Msun$ and 
$\SFR \approx 8\,\Msun\,\yr^{-1}$.
\citet{Lai07} employed near-IR data of the {\it Spitzer} IRAC to 
increase the reliability of stellar mass estimates, and derived 
$\Mstar=10^9 - 10^{10}\,\Msun$ and ages of $5-100$\,Myr using 3 LAEs 
at $z\sim 5.7$ in the GOODS northern field.  Their sample was the 3 out 
of 12 that had the IRAC detections, therefore they are likely to be 
the most massive ones.
In a more recent work, \citet{Ono10} produced stacked multiband images of 
165 ($z=5.7$) and 91 ($z=6.6$) LAEs, and found that these LAEs have 
low stellar masses of $(3-10)\times 10^7\,\Msun$, very young ages of 
$1-3$\,Myr, and negligible dust extinction through SED fitting.  
Overall, these observations suggest that LAEs are in general less massive,
have lower SFRs, and are younger than LBGs.

It has been suggested that LAEs might be exhibiting a very early phase of 
galaxy formation \citep[e.g.,][]{Hu96, Mori06b, Dijkstra07a, Kobayashi07}, 
where the \Lya\ photons emitted from the photoionized gas around star-forming 
regions are still able to escape from a relatively dust-free environment.  
However, it is also expected that \Lya\ emission might be observed from 
a later stage of galaxy formation \citep{Shapley03}, because the 
galactic outflows driven by SNe may evacuate the dusty gas around 
star-forming regions, as observed in local starburst galaxies 
\citep{Heckman01} and \highz\ LBGs \citep{Pettini01, Pettini02, Ade03}.
Recent studies by \citet{Finkelstein09b} and \citet{Pentericci09} revealed 
a population of LAEs with evolved stellar population, as well as a 
population of dusty starbursts, suggesting a wide variety of physical 
properties of LAEs. 

These ideas suggest evolutionary transitions from an early LAE to a LBG, 
and then back to a LAE at a later time.  If this scenario is correct, 
then LAEs could be short-lived phenomena that occur only at certain 
phases of galaxy formation, and we need to consider the duty cycle or 
stochasticity of LAEs, similarly to that of quasars 
\citep[e.g.,][]{Haiman01a}.

Furthermore, \citet{Kashikawa07} detected an interesting segregation
between LBGs and LAEs around a \highz\ quasar, and proposed  
that the existence of LAEs might be suppressed by the intense UV radiation
from nearby quasars.  This suggests that the existence of LAEs could  be
affected by various parameters of the local environment, such as overdensity
or intensity of the UV radiation field.

Another interesting development in recent observations is that the LF in the
rest-frame UV continuum {\em of LAEs} is constrained at the same time, 
as well as the \Lya\ LF.  This sets important additional constraints to
the evolution of LAEs from $z=6$ to 3, requiring the models to fit the UV LF
of LAEs as well as the \Lya\ LF.  The UV LFs of LBGs at $z=3-7$ are being
measured with increasing accuracy \citep[e.g.,][]{Ouchi04a, Yoshida06, 
Bouwens07, Iwata07, Bouwens09b, Oesch09, Ouchi09, Bouwens10a}, 
and the comparison between the UV LFs of LBGs and LAEs would give us 
information about the relationship between LBGs and LAEs, 
which is one of the key questions that is intensely studied in the field.

Motivated by these observations, there have been several theoretical studies
of LAEs using semianalytic models of galaxy formation.  \citet{Delliou05,
  Delliou06} used the semianalytic model of \citet{Baugh05} to show that the
\Lya\ LF of LAEs at $z=3$ can be explained if a uniform escape fraction of
\Lya\ photon $\fesc = 0.02$ and a top-heavy IMF is assumed.  
\citet{Dijkstra07a} proposed a model in which galaxies undergo a burst of 
very massive star formation that results in a large intrinsic \Lya\ equivalent 
width (EW), followed by a phase of Population II star formation with a lower EW.
\citet{Kobayashi07, Kobayashi10} used the semianalytic model of 
\citet{Nagashima05} with a standard IMF and introduced a variable escape 
fraction of \Lya\ photons owing to galactic wind feedback in \highz\ galaxies.

In this paper, we use cosmological hydrodynamic simulations of galaxy
formation based on the concordance $\Lam$ cold dark matter (CDM) model to
study the properties of LAEs, with a focus on the evolution of SFR, \Lya\ \&
UV LFs, stellar mass and clustering.  Previously, \citet{Nag04e} and
\citet{Night06} examined the properties of LBGs at $z=3-6$ using the same set
of cosmological simulations analyzed here, and showed that the rest-frame UV
LF of LBGs can be explained reasonably well with moderate extinction.
Building on top of our previous work on LBGs, here we discuss the relationship
between LBGs and LAEs.  

There have been other studies of LAEs using cosmological simulations
\citep[e.g.,][]{Furla05, Dayal09, Dayal10a, Dayal10b}. 
In particular, \citet{Dayal10c} have attempted to combine the results 
of radiative transfer calculations with LAE modeling, and examined
the \Lya\ and UV LFs of LAEs. 
Our approach here is somewhat different from theirs. We take a more 
minimal approach, and ask how much of the observed LAE properties can be 
explained by making the simplest assumptions on the escape fraction of 
\Lya\ photons or the stochasticity of LAE phenomenon. 

\begin{table*}
  \begin{center}
  \caption{Simulation Parameters}\label{table:sim}
    \begin{tabular}{lccccl}
      \hline \hline
Run & $L_{\rm box}$ & ${N_{\rm p}}$ & $m_{\rm DM}$ & $m_{\rm gas}$  & $\epsilon$ \\
\hline 
Q5  & 10.   & $2\times 324^3$ &  $2.12\times 10^6$ & $3.26\times 10^5$ & 1.23 \\
Q6  & 10.   & $2\times 486^3$ &  $6.29\times 10^5$ & $9.67\times 10^4$ & 0.82 \\
D5  & 33.75 & $2\times 324^3$ &  $8.15\times 10^7$ & $1.26\times 10^7$ & 4.17 \\
G6  & 100.0 & $2\times 486^3$ &  $6.29\times 10^8$ & $9.67\times 10^7$ & 5.33 \\
\hline \\
\multicolumn{6}{@{}l@{}}{\hbox to 0pt{\parbox{120mm}{\footnotesize 
Notes. Simulations employed in this study.  The initial number of gas particles is equal to that of dark matter particles, hence $\times 2$ for $N_{\rm P}$. The masses of dark matter and gas particles ($m_{\rm DM}$ and $m_{\rm gas}$) are given in units of $\himsun$, respectively, and $\epsilon$ is the comoving gravitational softening length in units of $\hikpc$.  The value of $\epsilon$ is a measure of spatial resolution. All runs adopt our ``strong'' wind feedback model. 
}\hss}}
\end{tabular}
\end{center}
\end{table*}

We have so far only discussed the point sources of \Lya\ emission, but there
are also sources with extended \Lya\ emission, called ``\Lya\ blobs''
\citep{Keel99, Steidel00, Matsuda04}.  One of the possibilities is that the
extended \Lya\ emission is powered by the release of gravitational potential
energy as the baryons condense inside dark matter halos \citep{Haiman00b,
Fardal01}.  
Using the same series of simulations as in this paper, \citet{Furla05} 
considered the \Lya\ emission from diffuse IGM and the gas in halos around 
galaxies, which is powered by gravitational processes and photoionizing 
background radiation.  They concluded that the \Lya\ emission from 
recombinations that follow the absorption of stellar ionizing photons 
(i.e., the \Lya\ emission associated with star formation) dominates the 
total \Lya\ photon production rate.  
Furthermore, \citet{Saito08} found that the number density of \Lya\ blobs
are only $10-20$\% of the total LAE population. 
Therefore, in this paper, we focus on the \Lya\ emission associated with 
star formation and do not consider \Lya\ blobs. 
We concentrate on the comparison with observations of LAE LF and their 
relationship to those of LBGs.

This paper is organized as follows.  In \S~\ref{sec:sim}, we briefly describe
our simulations.  We then discuss the specific SFR and stellar masses of
\highz\ galaxies in our simulations in \S~\ref{sec:ssfr}.  We present the
evolution of the galaxy stellar mass function in \S~\ref{sec:MF}, and discuss
the evolution of the \Lya\ LF from $z=6$ to 3 in \S~\ref{sec:LF}.  We propose
two simple scenarios to explain the \Lya\ LF of LAEs.  For both scenarios, we
compute the mean values of stellar mass (\S~\ref{sec:meanmass}), black hole
masses (\S~\ref{sec:bh}) and metallicity (\S~\ref{sec:metal}) of LAEs.  We
discuss the relationship between \Lya\ and UV LFs in \S~\ref{sec:UVLF}, and
their evolution in \S~\ref{sec:lf_evolve}.  The correlation function of LAEs
is presented in \S~\ref{sec:corr}, and the cosmic variance in the current 
surveys of LAEs is discussed in \S~\ref{sec:variance}.
We conclude in \S~\ref{sec:discussion}.


\section{Simulations}
\label{sec:sim}

We use the smoothed particle hydrodynamics (SPH) code {\small GADGET2}
\citep{Springel05e} in this work.  It employs the `entropy conserving'
formulation \citep{Springel02} to alleviate the overcooling problem, 
which previous generations of SPH codes experienced.  
Our simulations include radiative cooling by hydrogen and helium, 
heating by a uniform UV background \citep[e.g.,][]{Katz96a, Dave99}, 
star formation and supernova feedback based on a sub-particle 
multiphase ISM model \citep{Springel03b}, and a phenomenological model 
for galactic winds \citep{Springel03a}.

The details of the star formation model were described in \citet{Springel03b} 
and \citet{Nag04f}, so we only give a brief description here. 
In short, gas particles are allowed to
spawn a new star particle when a set of criteria is satisfied at each
time-step.  Groups of star particles are regarded as galaxies in the
simulation, and we identify them by applying the P-StarGroupFinder grouping 
algorithm developed by Springel. 
The star particles carry physical quantities such as their mass, formation
time, and metallicity. Using these tags, we compute the spectrum of each star
particle with the population synthesis code of \citet{BClib03}, and co-add the
individual luminosities to obtain the spectra of our simulated galaxies.
The inclusion of TP-AGB phase does not affect the results in this paper, as we are not dealing with the rest-frame near-IR flux of the simulated galaxies. 

We use four different simulations with varying box sizes and particle numbers
(see Table~1) in order to cover a wide range of halo masses and assess the
resolution effect. These simulations extend the set of runs carried out by
\citet{Springel03a} to higher resolution. Unfortunately, the Q6 simulation 
was stopped at $z\sim 4$ owing to its very long computing time. 
Therefore we basically use the Q5 run in this paper, 
and show the results from the Q6 run for $z=6$ where appropriate.
The results of the Q5 and Q6 runs are very similar at $z=6$, except that 
the Q6 run has a slightly better coverage for the lowest mass galaxies 
with $\Mstar \lesssim 10^{7.5}\,\Msun$.  The main conclusions of this paper 
are not affected by the absence of Q6 results at lower redshifts.  
The adopted cosmological parameters of all simulations considered here are
$(\Om,\Ol,\Ob,\sigma_8, h)= (0.3, 0.7, 0.04, 0.9, 0.7)$, where $h=H_0 /
(100\kms\,\mpc^{-1})$.

\begin{figure*}
\begin{center}
\FigureFile(160mm,360mm){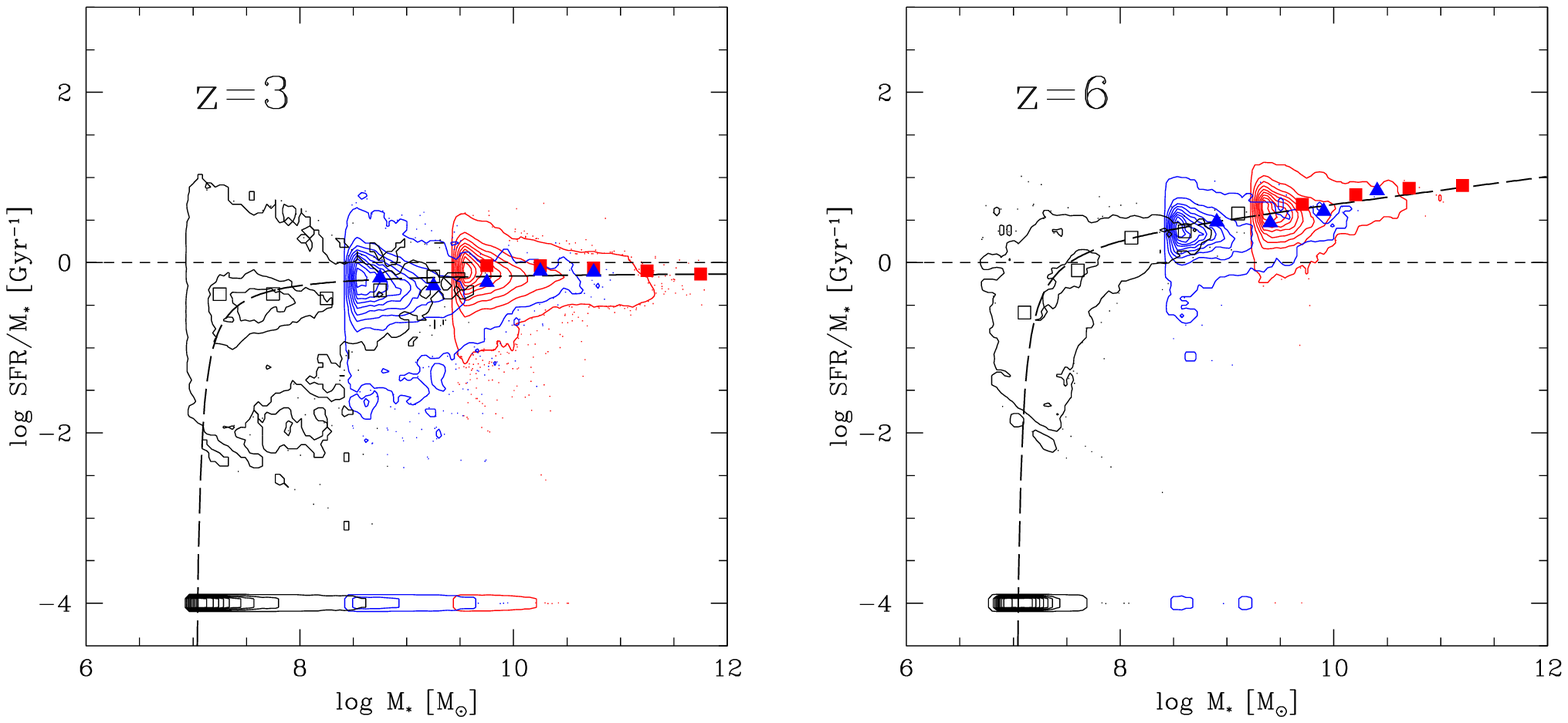}
\caption{Specific star formation rate vs. galaxy stellar mass at $z=3$ and
  6. The three sets of contours are for the Q5 (black), D5 (blue) and G6 (red)
  run from left to right. The median in each mass bin is shown with the
  symbols.  The analytic fits to the median points are shown with the
  long-dashed line and are given in the text.  Galaxies with zero SFR are
  indicated at $\log \SFR / M_\star = -4$.  }
\label{fig:ssfr}
\end{center}
\end{figure*}


\section{Specific SFR and Galaxy Stellar Mass}
\label{sec:ssfr}

We start by examining the instantaneous SFR of simulated galaxies at $z=3$ 
and 6, because we will use this quantity to calculate the \Lya\ luminosity.
Figure~\ref{fig:ssfr} shows the specific SFR ($\equiv \SFR / \Mstar$; SFR per
unit stellar mass) as a function of galaxy stellar mass at $z=3$ and 6.  The
three sets of contours are for the Q5 (black), D5 (blue) and G6 (red) runs,
from left to right.  The Q6 result at $z=6$ is very similar to that of Q5, so
it is not shown here.  Each simulation box can resolve only a limited range of
galaxy masses, so we use three different simulations to cover a wide range of
stellar masses, $\Mstar = 10^7 - 10^{12}\,\Msun$.  The median value of the
specific SFR is shown by the symbols for each bin of $\log\,\Mstar$. The
number of galaxies with no star formation is greater at $z=3$ than at $z=6$,
as indicated at $\log \SFR /M_\star = -4.0$.

The distribution broadens at the lower mass end of each contour for two
reasons. One is that there is a larger number of lower mass halos, therefore
the distribution naturally becomes broader as the larger population exhibits a
larger variation in its properties.  The other reason is that the resolution
limit of each run progressively shifts to lower masses, and close to the
resolution limit the distribution broadens owing to numerical noise.

At $z=3$, the specific SFR is almost constant (with significant scatter around
the mean) across the mass range of $\Mstar = 10^8 - 10^{11}\,\Msun$.  On the other
hand, at $z=6$, the specific SFR is an increasing function of galaxy stellar
mass, indicating that star formation is more efficient in more massive
galaxies at higher redshift. Star formation becomes rapidly inefficient 
in low mass galaxies with $\Mstar < 10^8 \Msun$, and the distribution seems 
to drop off completely at $\Mstar \simeq 10^7 \Msun$.  This rapid fall-off 
of the SFR at $\Mstar \simeq 10^7 \Msun$ may be related to the threshold 
density for star formation in the simulation and the observed SF cutoff in 
nearby spiral galaxies \citep{Kennicutt98}. 
A detailed investigation of the SF threshold in the simulation 
is beyond the scope of this paper, and is deferred to future work.

We provide the following approximate fit to the median points shown in
Fig.~\ref{fig:ssfr}:
\begin{equation}
Y = a X - \frac{b}{X-c} + d, 
\label{eq:sfrfit}
\end{equation}
where $(X,Y)=(\log\Mstar, \log \SFR / \Mstar)$, and $(a,b,c,d)=(0.0, 0.17,
7.0, -0.10)$ \& $(0.15, 0.2, 7.0, -0.75)$ for $z=3$ \& 6, respectively.  At
$z=3$, the down-turn at $\log\Mstar \approx 7.0$ is not clearly seen, but we
kept the value of $c$ the same for both redshifts for simplicity.

\begin{figure}
\begin{center}
\FigureFile(80mm,80mm){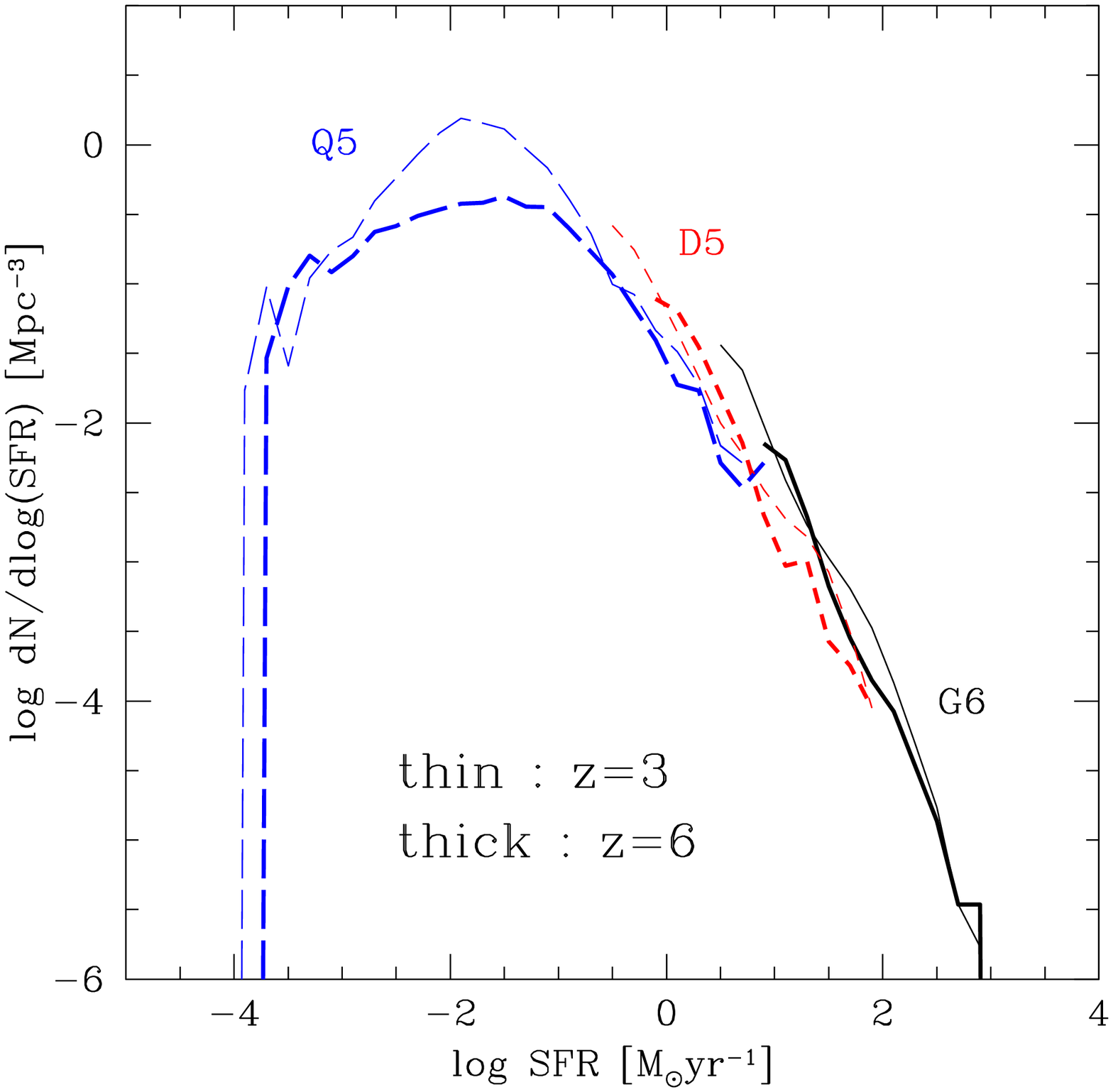}
\caption{SFR function at $z=3$ (thin lines) and $z=6$ (thick lines) for the Q5
  (long dashed), D5 (short dashed), and G6 (solid) runs.  }
\label{fig:sfrfunc}
\end{center}
\end{figure}

Figure~\ref{fig:sfrfunc} shows the galaxy SFR function at $z=3$ and 6, which
measures the differential number density of galaxies per logarithmic bin of
SFR, similarly to a galaxy luminosity function.  We find that there is not
much evolution in the SFR function from $z=6$ to 3.  This might seem
counter-intuitive given the evolution seen in the specific SFR, however, the
evolution of the SFR function is caused by a combined evolution in both the
specific SFR and galaxy stellar mass functions.

\section{Galaxy Stellar Mass Function}
\label{sec:MF}

\begin{figure}
\begin{center}
\FigureFile(80mm,80mm){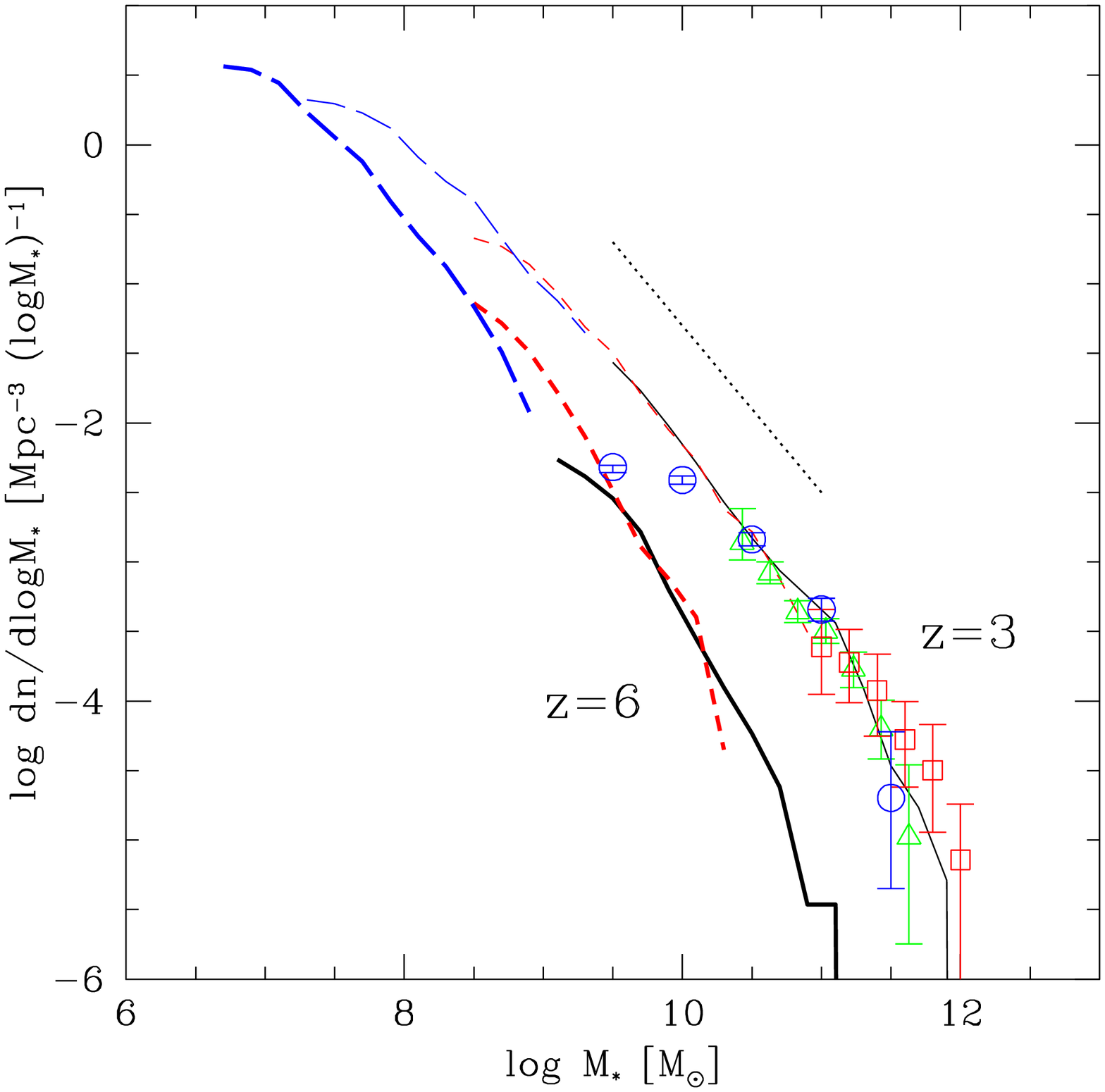}
\caption{Galaxy stellar mass function at $z=3$ (thin lines) and $z=6$ (thick
  lines) for Q5 (long dashed; Q6 for $z=6$), D5 (short dashed), and G6 (solid
  lines) runs.  The dotted line indicates the power-law of $n(\Mstar)\propto
  \Mstar^{-2.2}$.  The data points are from \citet[][blue circles]{Drory05},
  \citet[][green triangles]{Fontana06} and \citet[][red
  squares]{Perez08b}.  }
\label{fig:mf}
\end{center}
\end{figure}

In our simulations, the galaxy stellar mass function evolves rapidly from
$z=6$ to $z=3$, as expected in a hierarchical universe (Figure~\ref{fig:mf});
new halos of low mass are constantly formed, less massive galaxies merge into
more massive systems, and the number density of massive galaxies increases
with decreasing redshift. We note that galaxies grow in stellar mass in our
simulations, but the star formation becomes less efficient from $z=6$ to 3
(Figure~\ref{fig:ssfr}).  The latter effect compensates for the growth of the
mass function, resulting in little change in the SFR function
(Figure~\ref{fig:sfrfunc}).

At $z=3$, the simulation agrees well with the data from
\citet[][blue circles, $3.0<z<4.0$]{Drory05}, \citet[][green triangles, 
$2.5<z \leq 3.0$]{Fontana06} and \citet[][red squares, 
$3.0\leq z< 3.5$]{Perez08b} at $\log \Mstar \gtrsim 10.5$.  
The data by \citet{Drory05} suggest that the mass
function becomes shallower at $\log \Mstar < 10.0$, but our simulations have a
steeper slope ($n(\Mstar) \propto \Mstar^{-2.2}$) at the low-mass end as
indicated by the dotted line. While it appears likely that our simulations
overpredict the number of low-mass galaxies, future deeper observations are
needed to check this, based on a more reliable measurement of the faint-end of
the mass function. The location of the ``knee'' in the simulated mass function
is uncertain, given the steep faint-end slope.

We compute the total stellar mass density by integrating the interpolated mass
function in Figure~\ref{fig:mf} over the mass range of $\log \Mstar \approx
[7.0, 12.0]$, and obtain $\rho_\star = 3.5\times 10^8$ ($8.8\times 10^7$)
$\Msun\,\mpc^{-3}$ at $z=3$ (6).  This corresponds to $\Omega_{\star} =
0.0026$ ($6.5\times 10^{-4}$) at $z=3$ (6).  These values are higher than any
of the cosmic SFR models presented by \citet[][Figure~6a]{Nag06b}, suggesting
that the faint-end slope in Figure~\ref{fig:mf} might be too steep.
Nevertheless, we will use these total stellar mass densities in
\S~\ref{sec:meanmass} to compute the fraction of stellar mass that LAEs
contribute.

We also note that the stellar mass density predicted in our simulations 
is much higher than those of the current observational estimates
at $z=3$ and 6 \citep{Perez08b, Fontana06, Eyles07, Yan06b}. 
This is due to the large
number of low-mass galaxies in the simulation and the very steep 
faint-end of galaxy stellar mass function.  If one assumes the observed 
stellar mass density in the calculation, the fraction of 
stellar mass density contained in the currently observed LAEs 
would be much higher than the numbers presented in this paper.

\section{\Lya\ Luminosity function}
\label{sec:LF}

We compute the intrinsic \Lya\ luminosity $\La^{\rm int}$ emitted by \highz\
galaxies as
\begin{equation}
\La^{\rm int} = 10^{42}\,({\rm SFR} / \Msun\,{\rm yr^{-1}})~{\rm erg}\,{\rm s}^{-1},  
\label{eq:sfr}
\end{equation}
following \citet{Furla05}.  This relationship is accurate to within a factor
of a few according to the stellar population synthesis model of
\citet{Leitherer99} for a Salpeter initial mass function (IMF) with a mass
range of $1-100\,\Msun$ and metallicities between $0.05< Z/\Zsun <2$.  We use
the simulated SFR in the right-hand-side of Eq.~(\ref{eq:sfr}).

Without any corrections to the \Lya\ luminosity, we find that our simulations
overpredict the \Lya\ LF by a significant factor ($\sim 10$) compared to the
observations of \citet{Ouchi08}.  Here, we choose to compare our results with
the data by \citet{Ouchi08}, because their sample comes from a large survey
field and they have performed extensive comparisons with the earlier LF
estimates.  Ouchi et al.'s LF at $z\sim 3$ is consistent with that of
\citet{Gronwall07}.
According to Ouchi et al.'s data, there is not much evolution (no more than
a factor of $2-3$) between $z=6$ and 3 in the observed {\em apparent} \Lya\ LF
either in luminosity or number density.  However, \Lya\ fluxes from \highz\
sources are attenuated by the intergalactic neutral hydrogen, causing an
asymmetric profile in the \Lya\ emission line with the blue-side being
absorbed more \citep[e.g.,][]{Hu04, Kashikawa06b, Shimasaku06}.  Therefore,
when the data is corrected for this effect, little evolution in the {\em
  apparent} \Lya\ LF means strong evolution in the {\em intrinsic} LF, in the
sense that the \Lya\ luminosity and/or the number density of LAEs are
intrinsically brighter/higher at $z\simeq 6$ than at $z\simeq 3$.

In the following, we consider two possible scenarios to match the simulation
results to the observed {\em apparent} \Lya\ LF.  The two proposed scenarios
are very simple, but they capture the two extreme situations that plausibly
bracket the true behavior.


\subsection{Escape Fraction Scenario}
\label{sec:esc}

In the first scenario we simply assume that only a fixed fraction of \Lya\
photons reaches us from the source, i.e.,
\begin{equation}
L^{\rm obs}_{{\rm Ly}\alpha} = \fLya L^{\rm intrinsic}_{{\rm Ly}\alpha},  
\label{eq:Flya}
\end{equation}
where $L_{{\rm Ly}\alpha}$ is the \Lya\ luminosity.  The parameter $\fLya$ 
can be interpreted as an {\em effective escape fraction} that includes the 
following three effects: escape of ionizing photons, local dust extinction, 
and absorption by the IGM \citep{Barton04}.  We characterize this as
\begin{equation}
\fLya = \fdust\, (1-\fion)\, \figm,
\label{eq:esc}
\end{equation} 
where $\fdust$ is the fraction of \Lya\ photons that is {\em not} extinguished by
local dust, $\fion$ is the fraction of ionizing photons that escape from
galaxies and thus create no \Lya\ photons, and $\figm$ is the fraction of
\Lya\ photons that are {\em not} absorbed by the IGM, i.e., the transmitted
flux.  We call this case the ``{\em escape fraction}'' scenario.

Of course, in the real universe, different galaxies may have different values
of $\fdust$ and $\fion$, depending on their physical parameters such as age,
mass, SFR and local environment.  These parameters can also depend on
redshift.  We would need to perform radiative transfer calculations of 
\Lya\ photons to address these dependencies in detail.
Therefore, the above parameterization should be interpreted as an
attempt to capture the {\em average} behavior of bright galaxies that are
currently being observed, even though for simplicity we do not indicate the
implicit averaging with $\langle \cdots \rangle$ in our notation.

\begin{figure*}
\begin{center}
\FigureFile(160mm,160mm){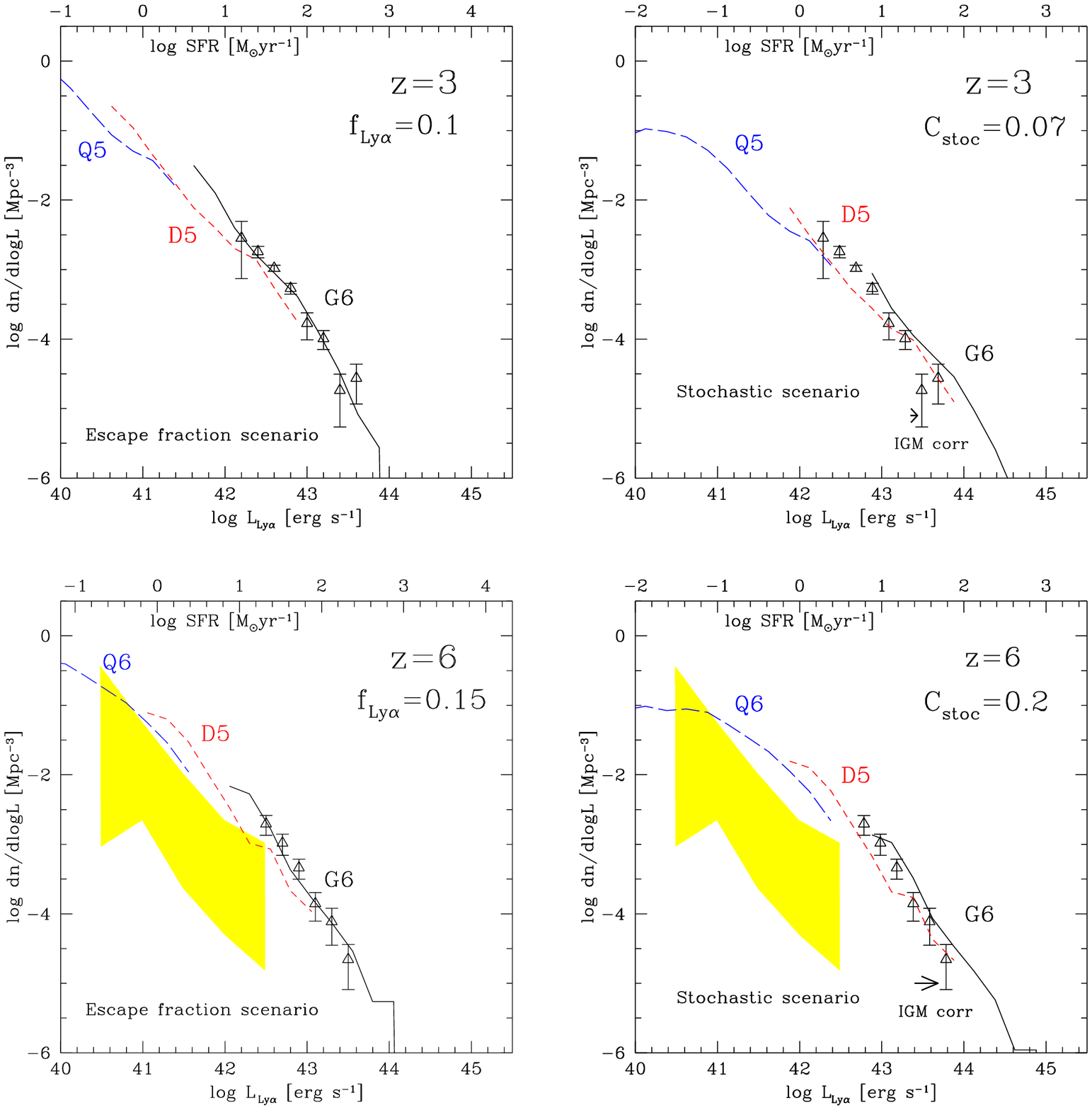}
\caption{ \Lya\ LF of LAEs at $z=3$ (top row) and $z=6$ (bottom row).  The
  data points are from \citet{Ouchi08} at $z=3.1$ and $z=5.7$.  The left
  column is for the ``{\em escape fraction}'' scenario, and the right column
  is for the ``{\em stochastic}'' scenario.  In the right column panels,
  corrections of $(\figm)^{-1}=(0.82)^{-1}$ \& $(0.52)^{-1}$ are applied to
  the data points of \citet{Ouchi08} at $z=3$ \& 6, respectively. The IGM
  attenuation as indicated by the small arrows.  The yellow shade in the
  bottom two panels indicates the region covered by the data points of
  \citet{Santos04a} and the simulation results of \citet{Dave06b}.  }
\label{fig:lf}
\end{center}
\end{figure*}

The left column of Figure~\ref{fig:lf} shows a comparison of our simulation
results with the observational data by \citet{Ouchi08}, adopting 
\begin{equation}
\fLya = 0.1\ (0.15)\ {\rm for}\ z=3\ (6).  
\end{equation}
This scenario corresponds to simply shifting the
simulated LF toward lower luminosity, therefore the currently observed LAEs
correspond to relatively massive galaxies with high SFR.  Here we selected the
values of $\fLya$ such that the G6 run agrees well with the observed data
points, because this run has the largest box size and covers the bright-end of
the observed LF much better than our other runs.  The D5 run underestimates
the number density of massive galaxies with $\log \La \gtrsim 42$ owing to its
smaller box size.  The agreement between the simulation results and the
observed data is very good at both $z=3$ and 6, including the slope of the LF.
Since our SFR function does not evolve very much (Figure~\ref{fig:sfrfunc}), the
values of $\fLya$ at $z=3$ and 6 are very close.

\vspace{0.5cm}

\subsubsection{IGM attenuation and $\fion$, $\fdust$}
\label{sec:IGM}

Many researchers \citep[e.g.,][]{Delliou06, Kobayashi07} simply adopted
$\figm=1.0$, arguing that various effects can reduce the amount of IGM
attenuation, such as ionization of the IGM around galaxies, clearing of the
IGM by galactic winds, and redshifting of \Lya\ photons by scattering in
the wind.  They also referred to the fact that the reionization of the Universe
was mostly completed by $z\sim 6$, as indicated by measurements of
Gunn-Peterson absorption in quasar spectra, and as suggested by
constraints on the clustering of LAEs \citep{Mcquinn07}.
Some observations also found that the \Lya\ lines are redshifted 
relative to the systemic velocity, suggesting strong outflows in 
high-$z$ LAEs \citep{Shapley03, McLinden10}.   
These facts suggest that the asymmetric \Lya\ line profile 
is likely to be caused by the absorption by the local ISM at the source, 
rather than by the IGM.  

However, on the other hand, $z\sim 6$ LAEs have asymmetric profiles 
with no blue emission, and have statistically less blue side emission 
compared to $z\sim 3$ LAEs \citep{Ouchi10}.  This fact suggests
that some IGM absorption does exist. Therefore, the true value 
of $\figm$ at $z=3$ is likely to be between the two cases that 
we will discuss below. 

Let us first consider the case of $\figm=1.0$.  
In this case, our result implies
\begin{equation}
\fdust\,(1-\fion) = 0.10\ (0.15)\quad {\rm for}\ z=3\ (6).  
\end{equation}
Adopting the values of $\fion=0.06$ (0.20) at $z=3$ (6) from 
\citet{Inoue06}, we obtain
\begin{equation}
\fdust = 0.11\ (0.19)\quad {\rm for}\ z=3\ (6). 
\end{equation}

\vspace{0.3cm}

Next, we consider the case in which the blue-side of \Lya\ emission 
is predominantly absorbed by IGM. 
We estimate the IGM attenuation factor to be
\begin{equation}
\figm = e^{-\tau_{\rm eff}} = 0.82\ (0.52)\ \;
{\rm for}\ z=3\ (6)
\label{eq:figm}
\end{equation}
using the \citet{Madau95} formulation with the assumption that only half of
the symmetric \Lya\ line is absorbed.  These values are consistent with those
obtained by \citet{Ouchi08}.
Inserting Eq.~(\ref{eq:figm}) into Eq.~(\ref{eq:esc}), 
we obtain 
\begin{equation}
\fdust\,(1-\fion) = 0.12\ (0.29)\quad {\rm for}\ z=3\ (6).
\end{equation}

\smallskip

\citet{Chen07} reported that, using the afterglow spectra of long-duration
gamma-ray bursts, the mean escape fraction of ionizing radiation from
sub-$L_*$ galaxies at $z\gtrsim 2$ is $\langle \fion \rangle = 0.02\pm 0.02$ 
with an upper limit of $\langle \fion \rangle \leq 0.075$.  
If the escape fraction of ionizing photons is as small as $\fion = 0.02$, 
then our result implies 
\begin{equation}
\fdust \approx
0.12\ (0.29)\ {\rm at}\ z=3\ (6).
\end{equation}

\citet[][Fig.~3]{Inoue06} compiled existing direct measurements of escape
fractions of ionizing photons and estimates based on the observed ionizing
background intensities.  They suggested that the value of $\fion$ might be
increasing with redshift: $\fion \approx 0.02, 0.06\ \&\ 0.2$ at $z=2, 3\ \&\
4-6$.  In this case,
\begin{equation}
\fdust = 0.13\ (0.36)\quad {\rm at}\ z=3\ (6). 
\end{equation}
The lower value of $\fdust$ at $z=3$ suggests that the environment around the
star-forming regions becomes more polluted by dust as star formation proceeds
from $z=6$ to 3, blocking more \Lya\ photons.

\smallskip

In either case, our results imply $\fdust \approx 0.1$ 
at $z=3$, and $\fdust \approx 0.2-0.4$ at $z=6$.

\vspace{0.5cm}

\subsection{Stochastic Scenario}
\label{sec:sto}

The other scenario we examine is based on the assumption that only a fixed
fraction ($\Csto$) of all star-forming galaxies can be observed as LAEs 
at a given time.
As far as the luminosity function is concerned, this is equivalent to assuming
that each LAE has a certain stochasticity and remains observable only for 
a limited duration of time, therefore, we call this case the 
``{\em stochastic}'' scenario.

In the case of $\figm = 1$, no corrections of the observed \Lya\ LF is
necessary. In this case, we obtain a good match between the simulated LF
and the observed one with 
\begin{equation}
\Csto = 0.06\quad {\rm for}\ z=3\ {\rm and}\ 6.
\end{equation} 

Next, let us consider the case where IGM absorption is important. 
In this case, we correct the observed \Lya\ LF for the effect of 
IGM attenuation by factors of $(\figm)^{-1} = (0.82)^{-1}$ 
\& $(0.52)^{-1}$ for $z=3$ \& 6, respectively, before we calculate 
the values of $\Csto$.  As shown in the right column of 
Figure~\ref{fig:lf}, we obtain good agreement with the data if we
assume 
\begin{equation}
\Csto = 0.07\ (0.2)\quad {\rm for}\ z=3\ (6).
\end{equation}

Both of these cases are equivalent to lowering the normalization of 
the simulated LF to match the observed data.
Here we adjust our simulated LF so that the results of the D5 and G6 runs
bracket the observed data points, because in this scenario the observed data
overlaps with lower mass galaxies in the D5 run.  The above value of $\Csto$
can be interpreted as either only 7\% of the sources are turned on as LAEs at
$z=3$, or LAEs are turned on only for 70 Myrs out of 1 Gyr at $z\approx 3$.

In principle, the value of $\Csto$ might depend on other physical 
quantities such as the age and metallicity of the galaxy.  
If we had a perfect simulation that could resolve all the physical 
phenomena including star formation and feedback on the smallest scales, 
then such dependencies between $\Csto$ and other physical quantities
should come out of our simulation naturally.  Unfortunately, 
current simulations are still limited in their scope, therefore some 
microphysics are missing. Those limitations are captured by 
this $\Csto$ parameter, and on the surface, it is simply the fraction 
of star-forming galaxies that would appear as LAEs in our simulation. 
Even with this simplest treatment, we are able to reproduce the \Lya\ LF
very well, and additional dependencies on age or other quantities 
are not necessary to match the observed \Lya\ LF. 
Most likely, the value of $\Csto$ will be strongly linked with the 
detailed star formation history and feedback processes. 
As we already described, part of the stochasticity due to star 
formation history is already taken care of within our simulation 
dynamically.  The remaining small scale physics that we cannot 
capture may cause additional stochasticity, and that is modeled with 
the $\Csto$ parameter in the present paper. 
In the future, when we carry out higher resolution simulations with 
more microphysics, we will start to be able to dissect the effect  
of $\Csto$ into different physical processes, and study its correlations
with different physical quantities. 

The yellow shaded region in the lower panels of Figure~\ref{fig:lf} indicates
the region covered by the data points of \citet{Santos04a} and the simulation
results of \citet{Dave06b}.  Their results are significantly lower than our
simulation results at $\log \La = [40.5, 42.5]$.  We comment further on this
point in \S\,\ref{sec:variance} and \S\,\ref{sec:discussion}.

In the remaining of the paper, we use the result of $\figm<1$ when 
necessary, with a cautionary remark that the true values lie somewhere
between $\figm = 0.82 - 1.0$ ($z=3$) and $0.52-1.0$ ($z=6$).


\section{Clarification of the models}

The formulation of the above two scenarios are primarily driven by the 
direction of the shift in the luminosity function (i.e., horizontal versus 
vertical shift) when trying to match the simulated LF with the observed one. 
The simulated LF (which is the starting point of our discussion) 
already takes account of the physics such as the number density of 
dark matter halos, radiative cooling of the gas, star formation, and 
supernovae feedback. 

However, one could take another approach in the spirit of the 
``halo occupation'' models, namely, by starting from the number density 
of dark matter halos and considering the physical processes that lead to 
the observed number density of LAEs \citep[e.g.,][]{Dijkstra07a}. 
In this case, the number density of observable LAEs can be written 
as follows:
\begin{equation}
n_{\rm LAE} = n_{\rm halo}\ f_{\rm SFthresh}\ f_{\rm SFduty}\ 
f_{\Omega}\  f_{\rm esc,\Omega}, 
\end{equation}
where $f_{\rm SFthresh}$ is the fraction of halos that ever have 
sufficiently high SFRs to generate enough \Lya\ photons to be 
observed as LAEs, $f_{\rm SFduty}$ is the duty cycle of SF activity 
of those halos at a given time, $f_{\Omega}$ is the fraction of solid angle 
that has non-zero \Lya\ photon escape, and $f_{\rm esc,\Omega}$ is the 
average escape fraction of \Lya\ photon within that solid angle. 
The term $f_{\rm esc,\Omega}$ can be further broken down into different 
physical processes as we did in Eq.~(\ref{eq:esc}).

Note that our simulations partially took care of the first two terms by 
simulating the gas infall onto dark matter halos and star formation 
processes dynamically.  Therefore the effects of $f_{\rm SFthresh}$ and 
$f_{\rm SFduty}$ are already imprinted on the simulated \Lya\ LF without 
any corrections, as represented by $L^{\rm intrinsic}_{{\rm Ly}\alpha}$
in Eq.~(\ref{eq:Flya}).

For the escape fraction scenario described in \S~\ref{sec:esc}, 
$f_{\Omega}=1$ is assumed before the IGM absorption, and the parameter $\fLya$ 
reflects only the effect of $f_{\rm esc,\Omega}$ and IGM absorption, i.e., 
$\fdust\, (1-\fion) = f_{\rm esc,\Omega}$.

For the stochastic scenario described in \S~\ref{sec:sto}, the parameter 
$f_{\Omega}$ is allowed to vary. 
This term is likely to be the major source of stochasticity in deciding
which halos are seen as LAEs from Earth, because \Lya\ radiative transfer 
is expected to be anisotropic and there may be a strong orientation effect.
Furthermore, it is possible that the combination of 
$f_{\rm SFthresh}\,f_{\rm SFduty}$ to vary as well, because our 
simulations are unable to follow the stochasticity of star formation 
activity below $\sim 100$\,parsec scales owing to limited resolution. 
If there are additional stochasticity in the SF activity, for example, due to 
a turbulence on small scales on top of the global gas dynamics followed
by our simulation, then there would be additional variation in 
$f_{\rm SFthresh}\,f_{\rm SFduty}$ on shorter time-scales than the simulation 
time-steps.  In summary 
$\Csto \approx f^{\prime}_{\rm SFthresh}\ f^{\prime}_{\rm SFduty}\ f_{\Omega}$, 
with a prime on the first two terms to indicate that they are on top of 
what is being simulated in our dynamical simulation. 
As we will later show in \S~\ref{sec:UVLF}, our rest-frame UV LF at $z=3$ 
agrees very well with the observations, therefore, the SFR and 
its stochasticity owing to the gas dynamics at $>$100\,pc scales is
simulated relatively well in our simulations. 

Which parameters and descriptions one would take is simply a matter of 
choice, and since our starting point was the simulated LF, we will use 
our original parameterization of $\fLya$ and $\Csto$ in the rest of this 
paper.  

\begin{figure*}
\begin{center}
\FigureFile(160mm,160mm){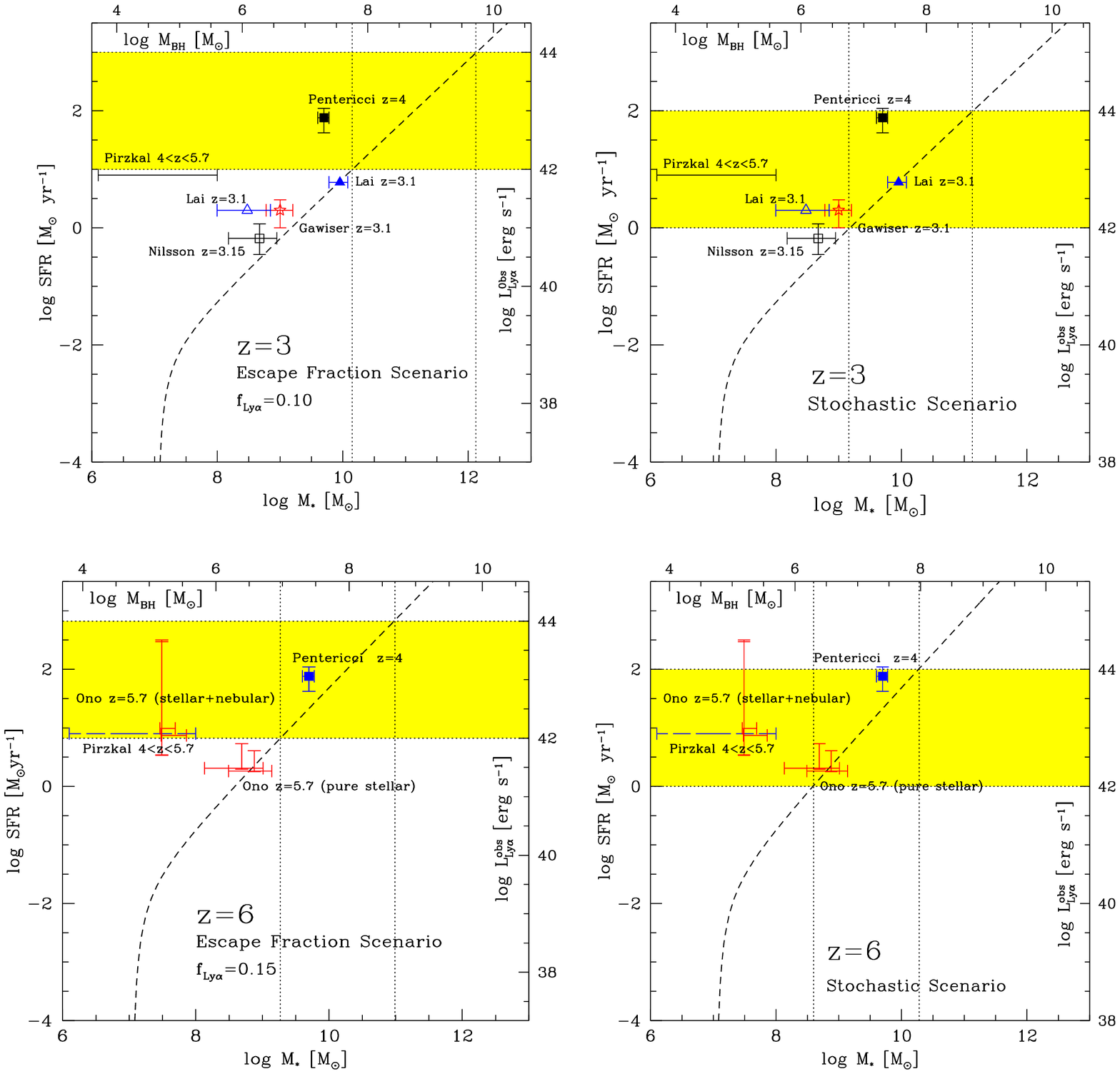}
\caption{Summary of relationships between galaxy stellar mass, SFR, \Lya\
  luminosity, and BH mass assuming a ratio of $\Mbh / \Mstar \approx 0.004$
  (0.005) at $z=3$ (6) \citep{Hopk07i}.  The dashed lines are based on
  Eq.~(\ref{eq:sfrfit}).  The yellow shaded regions indicate the currently
  observed luminosity range of $\log \La = [42, 44]$.  Data points are from
  \citet[][open star]{Gawiser07}, \citet[][open and filled triangles for
  IRAC-undetected and detected sample, respectively]{Lai08}, 
  \citet[][open square]{Nilsson07b}, \citet[][filled square]{Pentericci07}, 
  \citet[][long dashed horizontal bar in the bottom panels]{Pirzkal07}, 
  and \citet[][two sets of red error bars; one at lower $\Mstar$ is for the models including the effects of nebular emission lines, and the other at higher $\Mstar$ is using only stellar spectrum. Each set includes two error bars with metallicities of $Z/\Zsun=0.02$ and 0.2.]{Ono10}.   }
\label{fig:summary}
\end{center}
\end{figure*}


\section{Stellar mass of LAEs}
\label{sec:meanmass}

\begin{figure*}
\begin{center}
\FigureFile(160mm,80mm){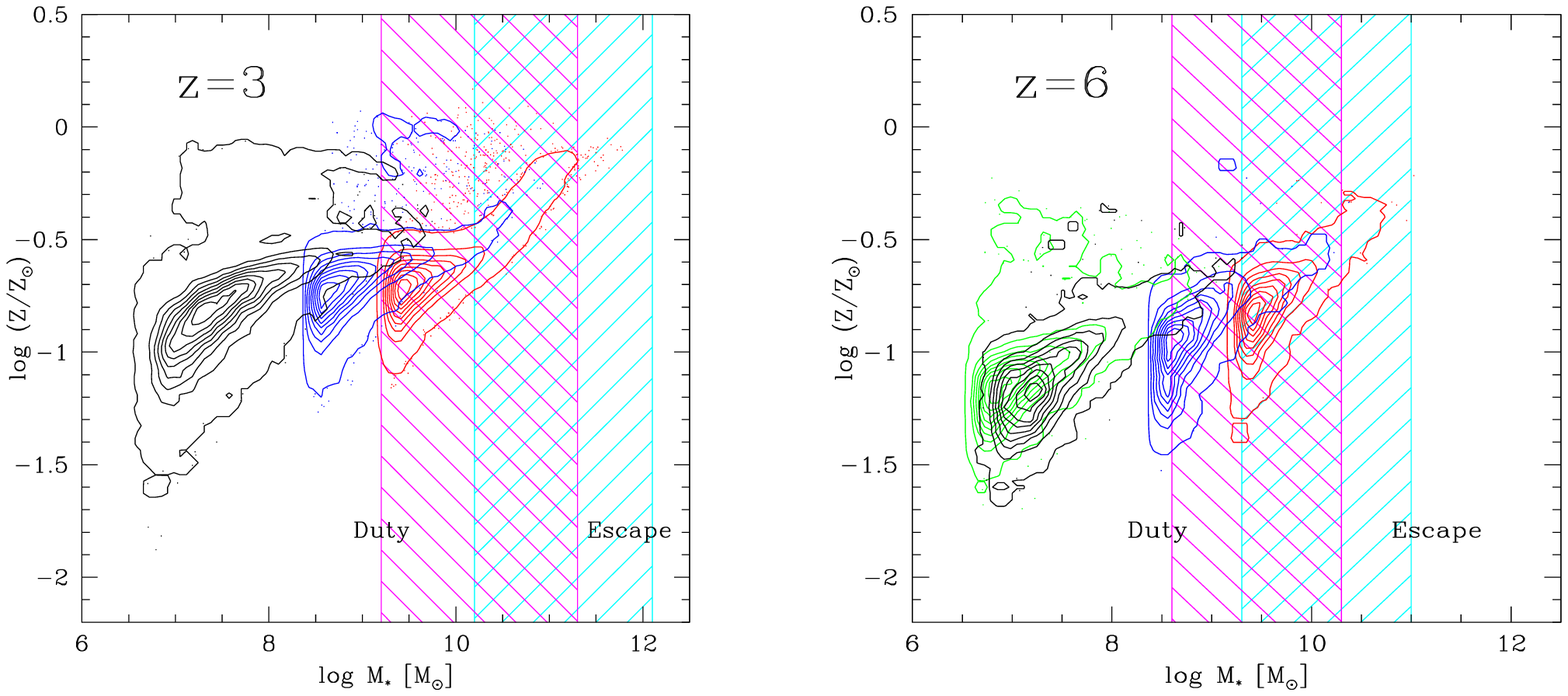}
\caption{Galaxy stellar metallicity vs. galaxy stellar mass.  The three sets
  of contours are for the Q5 (black), D5 (blue) and G6 (red) run from left to
  right.  For $z=6$, the result of the Q6 run is also shown with the green
  contour. The range of galaxy stellar mass for each scenario is indicated by
  the shade.  }
\label{fig:metal}
\end{center}
\end{figure*}

Figure~\ref{fig:summary} summarizes the relationship between galaxy stellar
mass, SFR and \Lya\ luminosity.  
The values of $\La$ being discussed in this section are the observed values
and the model predictions for observed values after IGM and any other
intrinsic absorption of \Lya\ photons.
The yellow shaded regions indicate the luminosity range of 
$\log \La = [42, 44]$, which roughly corresponds to the currently observed 
LAEs at $z=3-6$ \citep{Ouchi08}.  The corresponding stellar
mass ranges differ significantly, depending on the two scenarios and redshift
as summarized in Table~\ref{table:summary}, but they are not affected by the
uncertainties in the value of $\figm$, because Fig.~\ref{fig:summary} is
solely determined by Eq.~(\ref{eq:sfr}) and the values of $\fLya$.

In the escape fraction scenario, the raw (i.e., before any corrections)
simulated \Lya\ LF is simply shifted toward lower luminosity without a change
in the normalization, therefore currently observed LAEs correspond to the most
massive and luminous objects at the brightest end of the LF.  The mean stellar
masses of LAEs with $\log \La = [42, 44]$ in the G6 run are
\begin{equation}
\langle \Mstar \rangle = 2.5\times 10^{10}\ 
(1.9\times 10^9)\,\Msun\ {\rm at}\ z=3\ (6), 
\end{equation}
respectively, as given in Table~\ref{table:summary}.  These values are close
to the lower edge of the listed stellar mass ranges, because of the increasing
number of low-mass galaxies in a cold dark matter universe.  In this scenario,
the LAEs with $\log \La = [42, 44]$ contribute only 18\% (11\%) of the total
stellar mass density at $z=3$ (6).  This relatively low fraction owes to the
steep faint-end slope of the stellar mass function (Fig.~\ref{fig:mf}).

In the stochastic scenario, currently observed LAEs correspond to slightly
lower mass galaxies than in the escape fraction scenario:
\begin{equation}
\langle \Mstar \rangle = 3.8\times 10^9\ 
(6.1\times 10^8)\,\Msun\ {\rm at}\ z=3\ (6)
\end{equation}
for the LAEs with $\log \La =[42, 44]$ in the D5 run.  The fraction of stellar
mass density contributed by the LAEs is higher in this scenario, amounting to
42\% (29\%) for $z=3$ (6).

From this comparison only, we can see that more data points are covered
by the yellow shaded region, and the stochastic scenario is favored 
over the escape fraction scenario. 


\section{Black holes hosted by LAEs} 
\label{sec:bh}

The masses of black holes hosted by LAEs are also of significant interest, as
they determine the AGN contribution to the total energy output from LAEs.  We
indicate the black hole masses in the top axes of Figure~\ref{fig:summary}
assuming $\Mstar / \Mbh \approx 0.004$ (0.005) at $z=3$ (6), as suggested by
the recent numerical simulations of galaxy mergers
\citep{DiMatteo05,Robertson06b, Hopk07i} and observations
\citep[e.g.][]{Peng06}.  We find that the LAEs with $\log \La= [42, 44]$ host
BHs with masses $\log \Mbh = [7.7, 9.7]\ (z=3)$ and $[6.9, 8.7]\ (z=6)$ for
the escape fraction scenario, and $\log \Mbh = [6.7, 8.8]\ (z=3)$ and $[6.3,
8.0]\ (z=6)$ for the stochastic scenario.  If we instead assume the local
value of $\Mstar / \Mbh \approx 0.001$ \citep{Kormendy01}, then the above mass
ranges change to $\log \Mbh = [7.2, 9.2]\ (z=3)$ and $[6.3, 8.1]\ (z=6)$ for
the escape fraction scenario, and $\log \Mbh = [6.2, 8.3]\ (z=3)$ and $[5.7,
7.4]\ (z=6)$ for the stochastic scenario.

At $z=3.1$, \citet{Ouchi08} reported that the AGN fraction of LAEs is only 
$\sim 1$\% for LAEs with $\La > 1\times 10^{42}\,\erg\,\s^{-1}$, but that
the fraction is as high as 100\% for the bright LAEs with 
$\La > 4\times 10^{43}\,\erg\,\s^{-1}$.
On the other hand, they found no AGNs (nor LAEs) at $z=5.7$ with a bright 
luminosity of $\La > 4\times 10^{43}\,\erg\,\s^{-1}$, and suggested that 
the number density of LAEs with AGN activities would drop from $z=3$ to 6.  
Figure~\ref{fig:summary} shows that the mean BH mass for the LAEs with 
$\log \La = [42, 44]$ decreases from $z=3$ to 6, which is at least consistent
with the observed decrease of bright AGNs if BH mass is correlated with the 
strength of AGN activity.


\section{LAE Metallicity}
\label{sec:metal}

In Figure~\ref{fig:metal}, we show the galaxy stellar metallicity vs. stellar
mass in our simulations.  The metallicity of each galaxy is computed by
summing up the metal mass in all constituent star particles, and then dividing
by the total stellar mass of each galaxy.  There is a weak positive
correlation between the two quantities with significant scatter.

At $z=6$, the majority of galaxies have $\lgZ < -0.5$, and the most massive
ones with $\log \Mstar > 10.5$ have $\lgZ > -0.5$. There are some outliers
with $\lgZ > -0.5$ at the low-mass end of the distribution, which are small
galaxies that have just started to undergo star formation and
self-enrichment. Close to the resolution limit, the coarse sampling of the
galactic winds leads to substantial numerical scatter in the metal loss, which
in turn can temporarily produce high metallicities for some of the small
galaxies. However, the fraction of such outliers is very small with $\sim 6$\%
(0.2\%) of the total sample in the Q5 run at $z=3$ (6).

At $z=3$, the majority of galaxies still has $\lgZ < -0.5$, but the number of
galaxies with $\lgZ > -0.5$ has increased significantly since $z=6$, and the
most massive ones with $\log \Mstar > 11$ approach solar metallicity.

Since the galaxy stellar mass range is lower in the stochastic scenario than
in the escape fraction scenario, the corresponding galaxies have lower
metallicity in general, as summarized in Table~\ref{table:summary}.  Owing to
the large scatter in the distribution, the difference in the metallicity range
is not so large between the two scenarios, but the trend of lower metallicity
in the stochastic scenario is clearly seen in the mean metallicity values
listed in Table~\ref{table:summary}.  As expected, the mean metallicity at
$z=3$ is higher than at $z=6$ by about a factor of 2 in both scenarios.

It is useful to compare the mean metallicity of LAEs and LBGs.  The mean
metallicity of LBGs is known to be $\sim 1/3\, \Zsun$
\citep[e.g.,][]{Pettini04}.  At $z=3$, the mean metallicity of LAEs in the
escape fraction scenario is comparable to that of LBGs with
$\langle Z/\Zsun \rangle =0.39$, while it is lower for the stochastic scenario with
$\langle Z/\Zsun \rangle =0.21$.
The full discussion of $\Mstar - Z$ relationship in our simulations will be presented elsewhere. 

\begin{figure*}
\begin{center}
\FigureFile(160mm,80mm){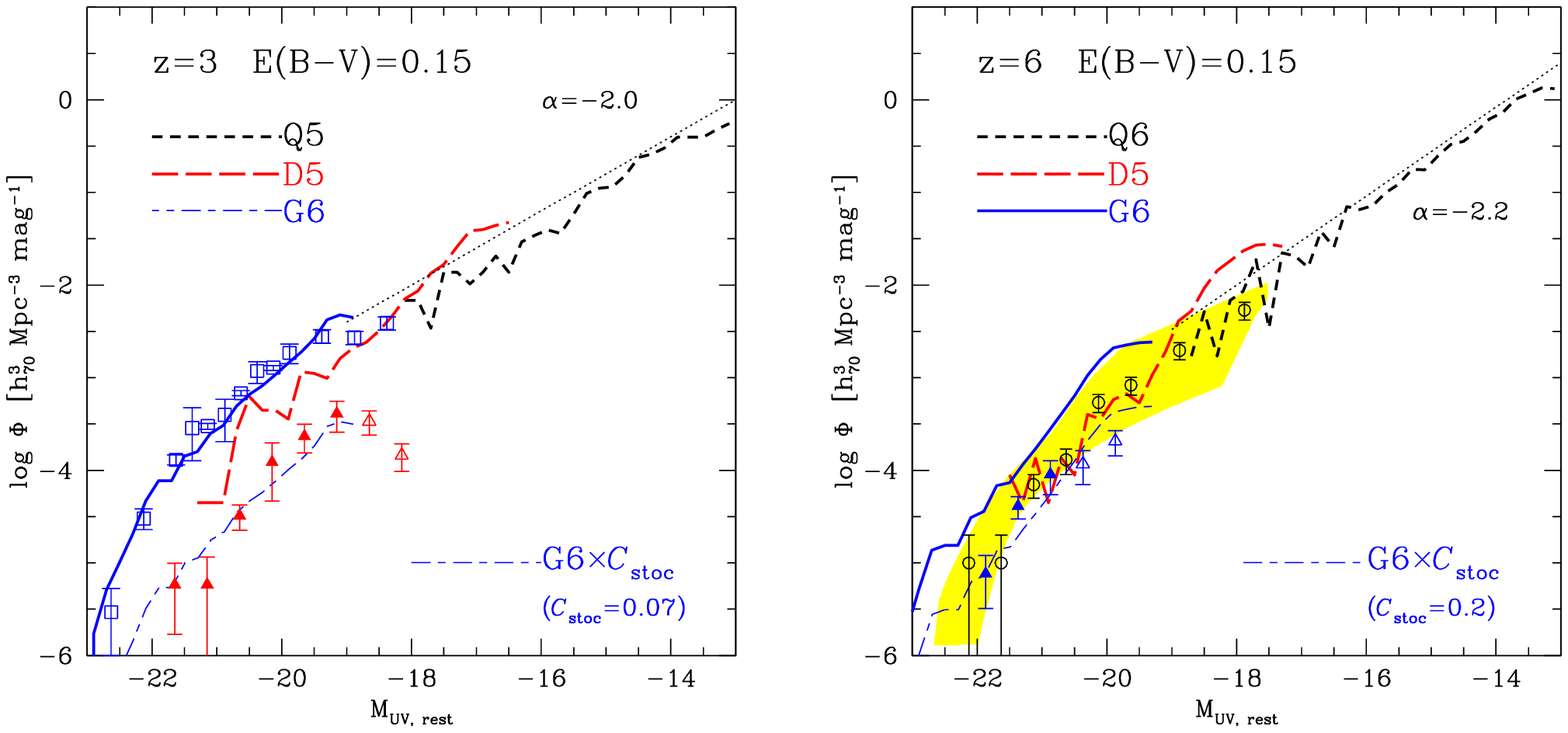}
\caption{Rest-frame UV luminosity functions at $z=3$ and $z=6$.  Thick lines
  are the original simulation results, and thin lines are those multiplied by
  $\Csto$. The data points are $z=3$ LBGs \citep[][blue open squares]{Ade00}, 
  UV LF of LAEs at $z=3.1$ \citep[][red filled triangles]{Ouchi08}, 
$i$-dropout LBGs at $z=6$ \citep[][black open circles]{Bouwens07} and UV LF 
of LAEs at $z=5.7$ \citep[][blue filled triangles]{Ouchi08}.  The two faintest 
data points of \citet{Ouchi08} indicated with open triangles are less reliable 
owing to incompleteness.  
The yellow shade encompasses the current observational estimates by 
\citet{Bouwens04a, Bunker04, Dickinson04, Yan04, Malhotra05, Beckwith06}.  
The black dotted line shows a power-law with a faint-end slope of 
$\alpha=-2.2$ in both panels with same normalization.  }
\label{fig:UV}
\end{center}
\end{figure*}


\section{Rest-frame UV Luminosity function}
\label{sec:UVLF}

One of the interesting recent development in the observations of LAEs is that
the rest-frame UV LF of LAEs is beginning to be constrained at the same
time as the \Lya\ LF.  This has been difficult in the past, because the UV
continuum of LAEs tends to be faint, and large samples of LAEs were not
available owing to limited sizes of the field-of-view (FoV) of the observations.

Figure~\ref{fig:UV} compares the simulated rest-frame UV LF with observational
data.  We adopt a uniform, moderate extinction of $E(B-V)=0.15$ at both $z=3$
and 6 following our previous work \citep{Nag04e, Night06}.  This extinction
value is the median value for the LBGs at $z\simeq 3$ \citep{Shapley01}.

At $z=3$, we obtain very good agreement between the G6 run (blue solid line)
and the observed data of \citet[][open squares]{Ade00} for LBGs.  This
agreement suggests that the SFR function (Fig.~\ref{fig:sfrfunc}) of simulated
galaxies at $z=3$ is quite reasonable at least at the bright-end.  We note
that it is also reasonable that the D5 run underpredicts the observed data owing
to its limited box size.

At $z=6$, we show in Fig.~\ref{fig:UV} the data by \citet{Bouwens07} for
$i$-dropout LBGs.  The yellow shade indicates the region covered by the
Schechter function fits of other observational studies \citep{Bouwens04a,
  Bunker04, Dickinson04, Yan04, Malhotra05, Beckwith06}, which we estimated
from Fig.\,11 of \citet{Bouwens07}.  The G6 run follows the upper envelope of
the yellow shade and slightly overpredicts the data at $M_{UV}<-22$.
\citet{Bouwens06a} reported that the extinction of $z=6$ LBGs are lower 
than those of $z=3$ ones, which would exacerbate the discrepancy between 
the simulation and observation at $z=6$. 

It might be possible to improve the agreement between the G6 run and the 
observed data by increasing the assumed extinction for massive galaxies.  
For example, correlating the extinction with metallicity could boost the 
extinction in massive galaxies (Fig.~\ref{fig:metal}), and make the 
bright-end of the LF steeper.  

Another possible cause for the overprediction of UV LF at the bright-end 
at $z=6$ is that the current simulations lack the explicit implementation 
of AGN feedback. It has been proposed that the energy and momentum 
feedback from supermassive black holes suppress the star formation in 
massive galaxies after going through the galaxy merger phase 
\citep[e.g.,][]{DiMatteo05, Springel05a}.
We plan to study the effects of AGN feedback on galaxy LFs at \highz\ 
in the future using hydrodynamic simulations that treat AGN 
feedback explicitly \citep[e.g.,][]{DiMatteo08, Sijacki07, Li07}.


\section{Evolution of \Lya\ and UV LF of LAEs}
\label{sec:lf_evolve}

Currently there is no clear picture regarding the evolution of the observed
rest-frame UV LF from $z=6$ to 3.  Some studies suggest a systematic
brightening of the characteristic magnitude $M^*_{\rm UV,rest}$ by
$0.5-1$\,mag \citep{Bouwens06a, Yoshida06, Bouwens07, Oesch09}.  Other work
argues for an evolution in the normalization $\phi^*$ \citep{Beckwith06} or in
the faint-end slope \citep{Yan04}.

In our current simulations, there is not much evolution in the UV LF from
$z=6$ to 3, as already shown by \citet{Night06}, except that the $\Muv^*$
becomes slightly brighter \citep[see Fig.~13 of][in which the comparison between
simulations and observed data was performed explicitly]{Bouwens07} and
$\alpha$ becoming slightly shallower, as can be seen in Fig.~\ref{fig:UV}
when compared to the power-law slope of $\alpha=-2.2$ (dotted line).  This
trend is at least qualitatively consistent with the observed one by
\citet{Bouwens07}.  The galaxy stellar mass function evolves as shown in
Fig.~\ref{fig:mf}, but the evolution in the SFR (Fig.~\ref{fig:sfrfunc})
cancels that out, resulting in little evolution in the UV LF.

Also shown in Fig.~\ref{fig:UV} are the UV LF {\em of LAEs} by
\citet[][filled triangles]{Ouchi08}.  Their data suggest that the number
density of LAEs at $z=3$ is only 10\% that of LBGs', down to $\Muv\simeq -20$
for the EW limit of $40-60$\,\AA, whereas the LAE fraction
at $z\sim 6$ increases to $50-100$\% for the EW limit of $\simeq 30$\,\AA.
Ouchi et~al.'s LAE fraction at $z=3$ is consistent with the earlier result by
\citet[][$\sim 25$\% with EW limit of 25\,\AA]{Shapley03}, but at $z\sim 6$, 
Ouchi et~al.'s LAE
fraction is higher than other spectroscopic studies of $i'$-dropout galaxies,
which typically suggest $\sim 30$\% \citep{Dow07, Stanway04b, Vanzella06}.

Let us first consider the stochastic scenario. 
At $z=3$, when we multiply the normalization of the simulated UV LF 
(which assumes uniform $E(B-V)=0.15$) by $\Csto=0.07$ (i.e., LAE fraction 
of 7\%), we obtain good agreement with the observed UV LF of LAEs.  
At $z=6$, we again obtain a reasonable agreement with the observed UV LF 
of LAEs when we multiply the LBG UV LF by $\Csto=0.2$ (i.e., LAE fraction of 
20\%), as shown in the right panel of Fig.~\ref{fig:UV} with the short-dash 
long-dashed line.  However, given that the LF of $i'$-dropout galaxies at 
$z\sim 6$ is still very uncertain and that our simulated LF might be 
overpredicting the UV LF at the bright-end, we cannot make a strong argument 
about the LAE fraction at this point.  If current observations are 
underestimating the number of very bright $i'$-dropouts ($\Muv \lesssim -22$) 
at $z=6$ for some reasons (e.g., cosmic variance, limited FoV), then our 
simulations would be able to explain both the UV and \Lya\ LF of LAEs at 
$z=3$ \& 6 with LAE fractions of 7\% and 20\%, respectively, in the 
stochastic scenario.

One possible problem in the above argument is that we implicitly assumed 
that the UV extinction of LAEs is also $E(B-V)=0.15$.  
Some observations suggested much lower extinction for LAEs with
$E(B-V)<0.05$ \citep{Ouchi08, Gronwall07, Ono10}, while others suggested 
a variety of stellar age, mass and dust properties 
\citep{Finkelstein09b, Pentericci09}.  

At this point there is no easy way out, unless we start considering
the possibility that the LAEs are a separate population from LBGs. 
It is certainly possible that the majority of LAEs are low-mass galaxies
with low extinction and low metallicity.  This is also consistent with 
what we have argued for in Fig.~\ref{fig:summary} for the stochastic scenario. 
However, by construction our approach assumes that both LAEs and LBGs are a single population of star-forming galaxies, and our current model cannot treat LAEs and LBGs as two separate population. We plan to address this issue in detail 
in the future by performing more direct radiative transfer calculations 
with dust models.


\section{Correlation function and bias of LAEs}
\label{sec:corr}

\begin{figure*}[t]
\begin{center}
\FigureFile(160mm,80mm){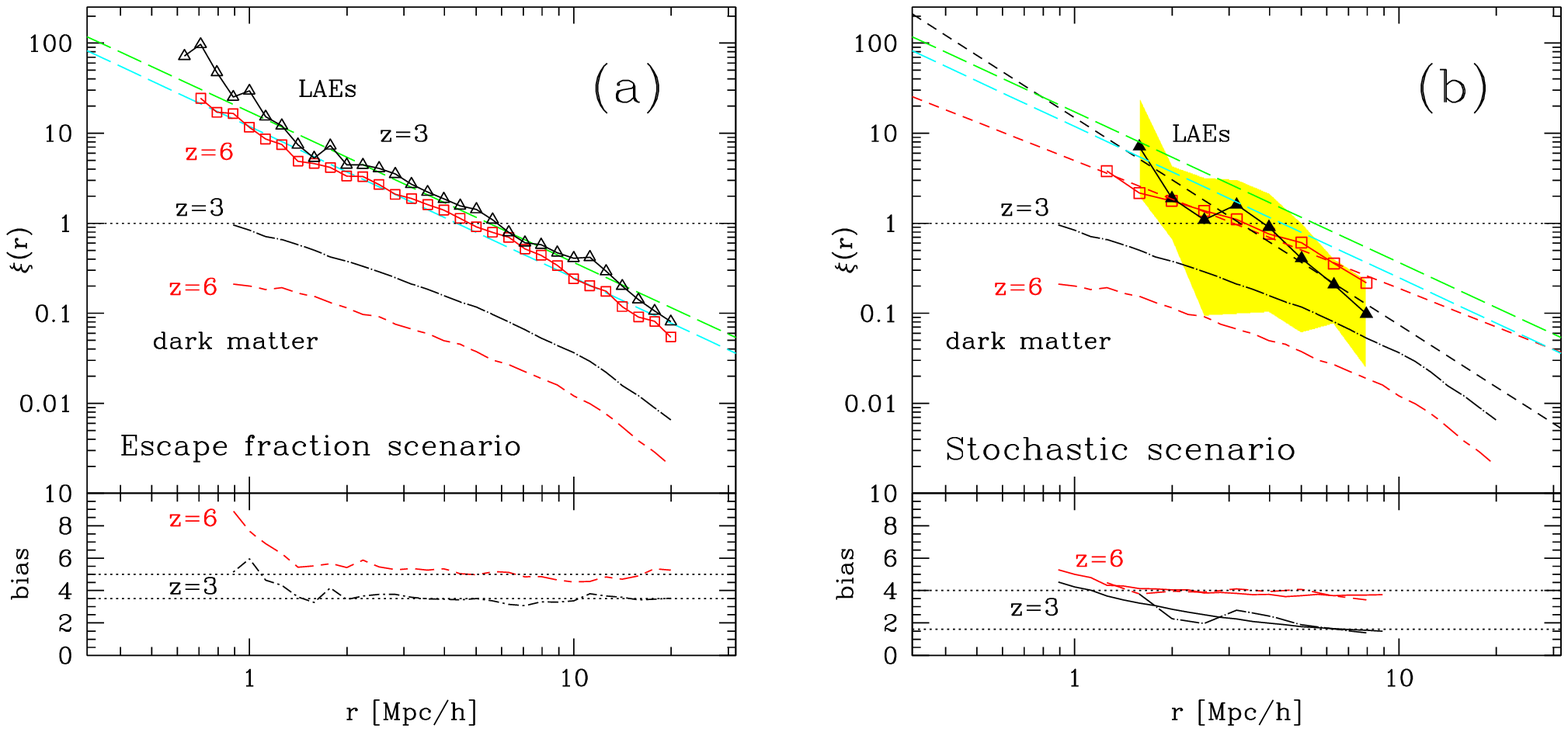}
\caption{Auto-correlation function of LAEs at $z=3$ \& 6. 
The G6 run was used for the escape fraction scenario (panel [$a$]), and the D5 
run was used for the stochastic scenario (panel [$b$]).  
The two {\em long-dashed} lines are the least square fits to the data points for 
the escape fraction scenario.  The parameters of the power-law fits are listed 
in Table~\ref{table:summary}.  In the stochastic scenario,
the sample size is small (101 \& 406 LAEs at $z=3$ \& 6), therefore we used 20 
different data sets to examine the variance of the CF (shown by the yellow shade 
for $z=3$).  The mean of the 20 trials is shown by the data points, and   
the power-law fits to the mean are shown in {\it short-dashed} lines.  
The variance at $z=6$ is smaller than at $z=3$ owing to larger sample.  
The bottom panels show the bias of LAEs against dark matter. 
For the stochastic scenario, the bias was computed for both 
using the direct simulation result ({\it short-long-dash} and {\it dot-dashed} 
lines) and using the power-law fits (solid lines). 
}
\label{fig:corr}
\end{center}
\end{figure*}

The correlation function of LAEs provides interesting constraints on the
distribution of LAEs, and it might help to discern the two scenarios that we
have discussed above.  We compute the auto-correlation function (CF) of LAEs
using the \citet{Landy93} estimator, $(DD-2DR+RR)/RR$, and show the results in
Figure~\ref{fig:corr}.  
Since the current luminosity limit of the observed LAEs is $\log
\La \approx 42.0$ for the $z=3$ sample, we restricted our sample to those with
$\log \La > 42.0$ for both $z=3$ \& 6 to measure the correlation function.  At
$z=6$, the current luminosity limit is higher at $\log \La > 42.5$ because of
the greater distances to the sources, but we kept the luminosity limit the
same for our calculations to obtain a reasonably strong correlation signal
with a sufficient number of simulated galaxies.

In the escape fraction scenario (Fig.~\ref{fig:corr}a), the observed 
luminosity range is completely
covered by the G6 run, so it is well-suited for this calculation.  Using the
data points in the range of $\log r = [0.0, 1.3]$, we perform a least square
fit to the power-law function $\xi(r) = (r_0/r)^{\gamma}$, and find $(r_0
[\himpc], \gamma)=(5.5, 1.67)$ and $(4.4, 1.68)$ for $z=3$ \& 6, respectively.
The CF at $z=3$ has a longer correlation length than at $z=6$, because the
mean stellar mass of LAEs at $z=3$ is higher (\S~\ref{sec:meanmass}). 
The slopes of the CF at the two epochs are very close to each other with 
$\gamma \approx 1.7$.  Our values of $(r_0, \gamma)$ are similar to those
of LBGs at $z\approx 3-6$ \citep{Ouchi04b, Ade05f}, which makes sense
because the mean stellar mass of LAEs in the escape fraction scenario is 
comparable to that of LBGs with a few times $10^{10}\,\Msun$ at $z=3$.

For the stochastic scenario, the observed data points of \citet{Ouchi08} are 
fully covered only by the D5 run (see Fig.~\ref{fig:lf}). The resolution limit of
the G6 run corresponds to $\log \La \approx 43$ in this scenario, therefore we
cannot use the G6 run to measure the CF for the stochastic scenario
with the same luminosity limit of $\log \La = 42.0$.  As discussed in
\S~\ref{sec:sto}, the number density of sources in the simulation has to be 
reduced by a factor of $\Csto=0.07$ (0.2) at $z=3$ (6) in this 
scenario.  
In the D5 run, there are $N_{>42} = 1439$ (2032) sources with $\log \La \ge 42.0$ 
at $z=3$ (6), therefore we need to select only $N_{>42} \times \Csto = 101$ 
(406) LAEs in the comoving volume of $(33.75\,\himpc)^3$.  
Owing to the small sample size, the CF signal for the stochastic scenario 
is somewhat noisy.  We therefore randomly resample 20 different data sets with 
above LAE numbers, and calculate the mean of the 20 different trials to reduce 
the noise.  We find that the CFs drop off at $r > 8\,\himpc$ owing to limited box 
size, therefore we only use data at $r < 8\,\himpc$ for the power-law fit.  
Fig.~\ref{fig:corr}b shows that the CF at $z=3$ is steeper with 
$\gamma = 2.3$, although we consider that this result is not reliable
owing to small sample size (101 LAEs). 
At $z=6$, we obtain a shallower slope of $\gamma = 1.49$ and $r_0=3.1\,\himpc$. 
In fact, if we increase the sample size at $z=3$ to the same size as that at $z=6$ 
(406 LAEs, corresponding to $\Csto=0.28$, which will overpredict the \Lya\ LF), 
then we obtain a similar signal to that at $z=6$ with $\gamma=1.61$ and 
$r_0=3.9\,\himpc$.  Therefore we consider that the steep slope of $z=3$ result
is simply owing to the limited sample size. 

The correlation length is smaller in the stochastic scenario than in the 
escape fraction scenario with $r_0 \simeq 3\,\himpc$, which is reasonable 
given the lower mean stellar mass of LAEs in this scenario.  
The sparse sampling also prohibits us from obtaining the CF signal at 
$r \lesssim 1.5\,\himpc$ in the stochastic scenario. 
In order to measure the CF for the stochastic scenario more reliably, we
need a simulation box size of $\Lbox \gtrsim 100\himpc$ with a resolution 
comparable to that of the D5 run.  This should become possible in the
near future thanks to rapidly increasing computing resources.


\subsection{Bias of LAEs}

The lower panels of Fig.~\ref{fig:corr} show the bias relative to the
clustering of the mass, which is defined as $b \equiv \sqrt{\xi_{\rm gal} / 
\xi_{\rm dm}}$.  We compute the correlation function of dark matter
by randomly sampling 200,000 dark matter particles in the G6 run.  
Using the D5 run yields very similar results on scales of $r = 1-8\,\himpc$. 
In both scenarios, the bias is a slowly decreasing function with increasing
distance. Even though the value of $r_0$ is greater at $z=3$, the bias 
relative to the 
dark matter is greater at $z=6$, because the growth in dark matter structure 
significantly increases the normalization of the dark matter CF from $z=6$ 
to 3.

In the escape fraction scenario (Fig.~\ref{fig:corr}a), we find 
$b\simeq 3.5$ (5.0) at $r=1.5-10\,\himpc$ for $z=3$ (6),   
crossing the above value at $r=4-5\,\himpc$.
At smaller scales ($r\lesssim 1.5\,\himpc$), the bias increases up to 
$b\sim 6$ (9) at $z=3$ (6).  This could owe to the excess clustering 
of galaxies on small scales as discovered by \citet{Ouchi05a}, 
although Ouchi's data at $z=4.0$ suggest that this increase in bias occurs 
at $r\lesssim 0.2\,\himpc$.  The increase of clustering on small-scales
can be ascribed to the substructures within each halo (``one-halo'' term).  
It is possible that the simulation is still lacking some physics or resolution 
to capture the correct scale for this transition. 
On large scales, the correlation function of dark matter seems to turn down, 
and at $z=6$ the bias somewhat increases at $r\gtrsim 15\,\himpc$, which is
probably a box-size effect.

In the stochastic scenario (Fig.~\ref{fig:corr}b), we use both the power-law 
fits and the actual simulation results to compute the bias. 
The calculation using the power-law yields monotonically declining functions 
with increasing distance. At $z=3$ the bias decreases from 4.5 to 1.5, and at 
$z=6$ from 5.3 to 3.7 with increasing distance. 
The calculation using the direct simulation CF shows more noisy behavior, 
wiggling around the power-law result.  At $z=3$ the bias decreases from 
3.8 to 1.4 at $r=1.6 - 8\,\himpc$.  At $z=6$, the wiggle is smaller and the 
{\it long-short-dashed} line 
agrees well with the power-law result, yielding
$b \simeq 4$ at $r=1.5 - 7\,\himpc$, which is contrasted with $b\simeq 5$ 
in the escape fraction scenario. 

The comparison to some of the observational estimates yields somewhat mixed 
results, but in general support the stochastic scenario. 
\citet{Kovac07} found $r_0 = 4.61\pm 0.6\,\himpc$ (taking the
contamination by randomly distributed objects into account) and $b\sim 3.7$
for the LAEs at $z\sim 4.5$ in the LALA survey \citep{Rhoads00}.
\citet{Ouchi03a} found $r_0 = 3.5\pm 0.3\,\himpc$ for $z=4.86$ LAEs.
\citet{Kovac07} pointed out that Ouchi's maximum permitted value would be 
$r_0 = 4.5\pm 0.4\,\himpc$ when the 20\% contamination by low-$z$ galaxies
\citep{Shimasaku04} is taken into account.  
Our results for the escape fraction scenario at $z=3$ \& 6 nicely bracket 
Kova{\v c} et~al.'s results at $z\simeq 4.5$.  The bias values in the duty 
cycle scenario brackets the Kova{\v c} et~al.'s result, but the value of 
$r_0$ is lower than theirs or at the lower edge of Ouchi et al.'s estimate. 

\citet{Gawiser07} reported $r_0=3.6^{+0.8}_{-1.0}$\,Mpc and 
$b=1.7^{+0.3}_{-0.4}$ for LAEs at $z=3.1$.  Our CF results at $z=3$ for the 
stochastic scenario are in good agreement with Gawiser et al.'s estimates. 
This is consistent with the nice agreement between our simulations
and the observed data seen in Fig.~\ref{fig:summary} for the stochastic 
scenario at $z=3$.  Our results from the escape fraction scenario at $z=3$ 
do not agree with Gawiser et al.'s data.  
At $z=6$, our result of $b=3.7$ at $r\sim 8\himpc$ in the stochastic scenario 
agrees well with $b=3.4\pm 1.8$ derived by \citet{Ouchi05b} for LAEs at 
$z\sim 5.7$. 
These comparisons again support the stochastic scenario.


\section{Cosmic Variance}
\label{sec:variance}

\begin{figure}
\begin{center}
\FigureFile(80mm,80mm){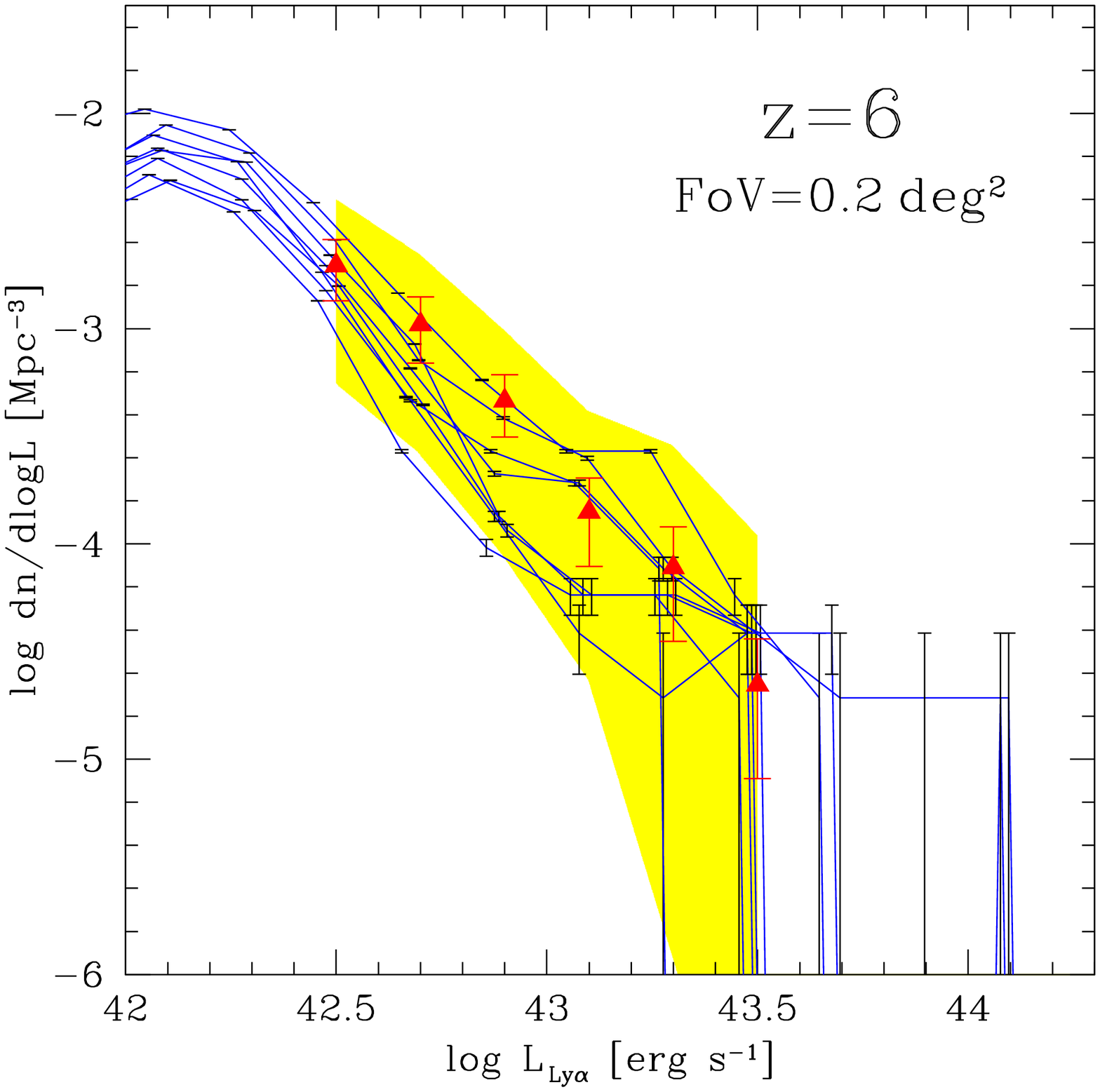}
\caption{\Lya\ luminosity function at $z=6$ measured in the eight subvolumes
of  $(45 \times 45\times 44\,\himpc)^3$ in the G6 run, corresponding to 
a $0.2\,\deg^2$ field of view.  The error bars show Poisson errors in each 
subvolume, slightly offset from each other to avoid overlap. 
The red filled triangle data points are from \citet{Ouchi08}, 
and the yellow shade shows the variance of their five $0.2\,\deg^2$ fields. 
}
\label{fig:lf_subvol}
\end{center}
\end{figure}

It is clearly difficult to conduct a deep, wide-field survey of \highz\
galaxies and obtain a statistically representative sample of galaxies in a
large volume of space. If the survey area is too small, then the observed
sample may not be representative of the total population owing to the
large-scale structure of the Universe, and the estimated LF would scatter
around the true LF.  This is one manifestation of the so-called ``cosmic
variance''.

In order to estimate the cosmic variance in our results, we use 
eight sub-volumes of $(45 \times 45\times 44\,\himpc)^3$ in the G6 run and 
derive the LF from each sub-volume (Figure~\ref{fig:lf_subvol}).  
For our adopted flat $\Lam$ cosmology, a
viewing angle of 1\,degree corresponds to about comoving $100\,\himpc$ at
$z=6$, so the above subvolume corresponds to a field with ${\rm FoV} = 
(0.45\,\deg)^2 \approx 0.2\,\deg^2$.  The thickness of $44\,\himpc$
was chosen to match the data of \citet[][Fig.~18]{Ouchi08}, 
but the exact value is not so important, as we would obtain a similar
result even if we adopted $(50\,\himpc)^3$ subvolumes.  
Here we used $\fLya = 0.15$ as we did in Fig.~\ref{fig:lf}. 

The large scatter of LFs seen in Fig.~\ref{fig:lf_subvol} owes to 
cosmic variance, and it clearly exceeds Poisson errors shown by the 
error bars.   
The red filled triangle is the data by \citet{Ouchi08} and the 
yellow shade is the variance of their data from five $0.2\,\deg^2$ fields. 
The cosmic variance we find is consistent with the field-to-field variation
observed by \citet[][Fig.~18]{Ouchi08}, as well as the data of 
\citet{Shimasaku04} that shows very different distribution of LAEs at 
$z=4.79$ and 4.86, separated by $\sim 40\,\himpc$ in the same FoV
\citep[see also][]{Hu06}.
At $\log \La=43.0$, the cosmic variance in the vertical direction is almost 
an order of magnitude in ${\rm d}n/{\rm d}\log \La$, and $\sim 0.6$\,dex 
in the horizontal direction (i.e., $\log \La$).
At $\log \La > 43.5$, the Poisson error bar is large, because one object in 
the above subvolume corresponds to the data point of 
$\log ( {\rm d}n/{\rm d}\log \La ) = -4.72^{+0.30}_{-\infty}$. 
Our result shows that a survey field of $\gtrsim 1\,\deg^2$ is necessary 
to obtain a reliable estimate of \Lya\ \& UV LFs at $z=6$. 

At $z=3$, the equivalent plot to Fig.~\ref{fig:lf_subvol} would have
an even greater scatter than at $z=6$.  Therefore a FoV of $3-4\,\deg^2$ 
would be desirable for a reliable estimate of LF. 


\section{Conclusions \& Discussions}
\label{sec:discussion}

Encouraged by the earlier successes in describing the properties of LBGs at
$z=3-6$ using cosmological SPH simulations \citep{Nag04e, Night06}, we
considered two simple scenarios to explain the luminosity functions of LAEs at
$z=3$ and 6.  These scenarios are very simple, but should capture the two
extreme cases of LAE properties.  The true physical nature of LAEs could be a
combination of the two scenarios.  We summarize the parameters associated with
the two scenarios in Table~\ref{table:summary}.  The main conclusions of our
work can be summarized as follows:
\begin{itemize}
\item In our simulations, star formation becomes progressively less efficient
  from $z=6$ to 3, especially for the massive galaxies with $\Mstar >
  10^{10}\,\Msun$.  This evolution in the SF efficiency largely compensates
  the growth in the galaxy mass function, yielding little evolution in the
  rest-frame UV LF.  We provide a fitting formula for the SFR--$\Mstar$
  relationship in Eq.~(\ref{eq:sfrfit}).  The SFR function
  (Fig.~\ref{fig:sfrfunc}) does not evolve very much from $z=6$ to 3.
\item We investigate two simple scenarios to characterize the properties of LAEs:
  the ``{\em escape fraction}'' scenario and the ``{\em stochastic}''
  scenario.  By matching the simulated \Lya\ LF of LAEs to the observed one by
  \citet{Ouchi08}, we find that the {\em effective} escape fraction of \Lya\
  photons is $\fLya=0.1$ (0.15) for $z=3$ (6), including the effect of IGM
  attenuation.  In the case of stochastic scenario, we find
  that the stochasticity parameter is $\Csto = 0.07$ (0.2) for $z=3$ (6), after
  correcting the observed LF for the IGM attenuation effect.  If we do not
  correct for IGM attenuation (i.e., $\figm=1.0$), 
we obtain $\Csto =0.06$ for both $z=3$ and 6.
\item We note that the value of $\figm$ is rather uncertain, so the 
most reasonable assumption is that its value is somewhere in-between 
$\figm = 0.82 - 1.0$ for $z=3$, and $\figm = 0.52 - 1.0$ at $z=6$. 
Therefore the other parameters discussed in this paper may also take 
correspondingly wide ranges of values depending on the IGM attenuation 
correction. 
\item In both scenarios, we find $\fdust \approx 0.1$ at $z=3$ and 
$\fdust \approx 0.2-0.4$ at $z=6$. In detail, if we correct for the IGM
attenuation by $\figm = 0.82$ (0.52) and assume $\fion = 0.06$ (0.2) 
at $z=3$ (6), then we obtain $\fdust = 0.13$ (0.36) at $z=3$ (6), 
respectively.  If we instead adopt $\figm = 1.0$ and the same values of 
$\fion$, then we find $\fdust = 0.11$ (0.19) for $z=3$ (6). 
\item The mean stellar mass of LAEs is lower in the stochastic scenario than
  in the escape fraction scenario, as summarized in Table~\ref{table:summary}.
  This implies lower mean values of BH mass, SFR and metallicity, as well as a
  lower amplitude of the correlation function in the stochastic scenario.  The
  mean stellar mass of LAEs is higher at $z=3$ than at $z=6$, as expected in
  the hierarchical structure formation model.
\item In the escape fraction scenario, the auto-correlation function of LAEs
  is similar to that of LBGs with $r_0 = 4-6\,\himpc$ and $\gamma \simeq 1.7$
  for both $z=3$ and 6.  The corresponding galaxy--dark matter bias is 
$b \simeq 3.5$ (5.0) at $r=1.5-10\,\himpc$ for $z=3$ (6).  These results bracket 
the observational estimates by \citet{Kovac07} at $z\sim 4.5$. 
On the other hand, the CF in the stochastic scenario has a shorter correlation 
length of $r_0 \simeq 3\,\himpc$ and the bias of $b=1.6-4.6$ ($z=3$) \& 4 
($z=6$), in good agreement with the observational estimates by 
\citet{Gawiser07} at $z\sim 3$.  In both scenarios, the bias parameter 
decreases with increasing distance. 
\item We find that the effect of cosmic variance on LF estimates can be
  quite strong.  The \Lya\ LFs at $z=6$ measured in 8 different fields of 
$\sim 0.2\,\deg^2$ have a variance of $\sim 1.0$\,dex in ${\rm
    d}n/{\rm d}\log \La$ at $\log \La=43.0$, and $\sim 0.6$\,dex in the
  horizontal direction (i.e., $\log \La$).  This result suggests that a survey
  field of $\gtrsim (1\,\deg)^2$ is necessary to measure the \Lya\ and
  UV LFs at $z=6$ reliably.
\item We find that the stochastic scenario is preferred over the escape 
fraction scenario through the comparisons with various observational data,
including $\Mstar$, SFR, \Lya\ LF, UV LF, correlation functions, and bias
relative to the dark matter distribution.    
In particular, the stochastic scenario succeeds in explaining the the rest-frame 
UV LFs of {\em both} LBGs and LAEs at $z=3$ \& 6, provided 
$E(B-V)\approx 0.15$ for both population. 
However, if the extinction of UV photons in LAEs is systematically lower than 
that of LBGs with $E(B-V)<0.05$ as the recent observations suggest 
\citep{Ouchi08, Gronwall07, Ono10}, then our simple models with {\it uniform} 
$\fLya$ or $\Csto$ fail to explain the UV LFs of LAEs and LBGs simultaneously.
In this case, the escape fraction scenario completely breaks down, because
it would predict a brighter UV LF of LAEs at $z=3$, contrary to the observation. 
\end{itemize}

In the case of stochastic scenario, additional modeling would be 
necessary to accommodate lower extinction for LAEs, such as variable
extinction and treating LBGs and LAEs as separate populations. 
The assumptions of linear SFR -- $\La$ relationship (Eq.~\ref{eq:sfr}) 
and uniform $\fdust$ \& $E(B-V)$ imply that galaxies with higher SFR 
would appear as brighter LBGs with higher $\La$.  
If we further assume that $E(B-V)$ is positively correlated with 
metallicity, then the mass-metallicity relationship implies that 
the brighter LBGs would be relatively massive galaxies with higher 
extinction.  Combined, this means that higher $\La$ sources have 
higher extinction, which is the opposite of what is found observationally
\citep{Shapley03, Gronwall07, Gawiser07, Ouchi08, Ono10}. 
One may consider that the sources with high extinction would have less observable \Lya\ flux due to a very strong extinction effect.  However within the framework of stochastic scenario, such sources are considered as inactive LAEs, and such an effect is already taken into account by the $\Csto$ parameter.  Those that are observed as LAEs should follow the correlations described above within 
our assumed model. 
Therefore, to assign lower extinction values for LAEs compared to the LBGs,  
one must invoke some other physical processes, such as galactic outflows, 
that could temporarily reverse the above relationship between $\La$ and $E(B-V)$. 

\vspace{0.2cm}

If the stochastic scenario is really the correct one, then it could explain
both \Lya\ and UV LFs of LAEs as follows: in the universe as early as $z=6$,
LBGs are actively forming stars, and as much as 20\% ($\Csto=0.2$) of them
would appear as LAEs (right panel of Fig.~\ref{fig:UV}).  If our simulation
is overpredicting the UV LF of LBGs at $z=6$, then this fraction could 
increase even more. 
By the time the universe has evolved to $z=3$, the star formation
becomes less efficient owing to the lower mean density and less supply of
infalling gas than at $z=6$.  The extinction of \Lya\ photons by dust also
becomes stronger at $z=3$ ($\fdust \sim 0.1$).  Only a small fraction
($\Csto=0.07$) of LBGs now appear as LAEs at $z=3$.  
Our values of $\Csto$ nicely
bracket the result obtained by \citet[][$\Csto=0.075-0.15$ at
$z=4.5$]{Malhotra02}.  If the star formation time-scale for the $z=3$ LBGs is
$\sim 200$\,Myr, then the life-time of LAEs at $z=3$ would be $\sim 14$\,Myr
according to the above stochasticity parameter.  Ages between several Myr to
100\,Myr are supported by recent observations \citep{Gawiser06, Lai07,
  Lai08, Pirzkal07}.  We remark that sporadic star formation histories are
automatically included in our results, since they occur naturally in our
dynamic simulations \citep{Nag05a}.  Therefore, the values of $\Csto$ 
in this paper characterize the stochasticity {\em on top of} the intermittency 
of the SF activity.

\vspace{0.2cm}

Because we adopted a single relationship between SFR and \Lya\ luminosity 
(Eq.~\ref{eq:sfr}), we are essentially assuming that the {\it intrinsic} 
equivalent width (EW) distribution of the \Lya\ emission line is a 
$\delta$-function at $\sim 70$\,\AA.\footnote{
We calculated the EW as follows: ${\rm EW} = \La / L_{\lam}(1216\AA) = 
\La / ([c/\lam^2] L_{\nu}[1216\AA]) = 70\,\AA$, where we used 
$\La = 10^{42}\,\erg\ \s^{-1}$ per SFR from Eq.~(\ref{eq:sfr}), and 
$L_{\nu}(1216\AA) = L_{\nu}(1500\AA) = 7\times 10^{27}\,\erg\ \s^{-1}\,{\rm Hz}^{-1}$ per SFR for a Salpeter IMF with mass limits $[0.1, 100]\Msun$ 
\citep{Madau98, Kennicutt98} assuming that the continuum spectrum of 
starburst galaxies is flat at UV wavelengths. 
}  
This is obviously oversimplified, given that the actually observed data has 
a fairly wide distribution \citep[e.g., Fig.~23 of][]{Ouchi08}.  
The effect of IGM absorption alone reduces the EW, and any stochasticity
in the escape fraction will broaden the EW distribution.
One can also accommodate a model that assumes a certain distribution of 
EW by, e.g., linking it to the scatter in the dust and metallicity 
distribution, and observe its effect on the LF.  
\citet{Kobayashi07} performed such an exercise using a semi-analytic
model of galaxy formation, and showed that the bright-end of the \Lya\ LF can
be described better with such a treatment.  Currently it is not clear whether
our simulation is overpredicting the \Lya\ LF at the bright-end, because 
there are no reliable
data at $\log \La >43.5$ (see Fig.~\ref{fig:lf}).  If future
observations show that there is a sharp exponential cutoff at $\log \La
>43.5$, we would be able to make stronger arguments on the possible
connections between feedback effects and the bright-end of the \Lya\ LF.  We
note that our current simulations already include the effect of SN feedback
and galactic outflows as described in \S~\ref{sec:sim}.  Given these
uncertainties, we decided not to invoke an additional model for the EW
distribution on top of what has been calculated in the simulation.

\vspace{0.2cm}

\citet{Dave06b} computed the \Lya\ LF using cosmological SPH simulation in a
similar fashion to our present work.  They assumed \Lya\ emission of
$2.44\times 10^{42}\,\erg\,\s^{-1}$ per SFR [$\Msun\,\yr^{-1}$]
\citep{Schaerer03} and took the metallicity variation into account.  They
argued that they obtain good agreement with the data of \citet{Santos04a} at
$\log \La =[40.5, 42.5]$ if they assume $\fLya=0.02$.  However, the LF data
points of \citet{Santos04a} are based on only a few objects and their
normalization seems to be too low to smoothly connect with the data points of
\citet{Ouchi08} at $\log \La > 42$, as shown by the yellow shade in
Fig.~\ref{fig:lf}.  Our simulations suggest that the LF at $\La =[40.5, 42.5]$
would be higher by a factor of $\sim 5$ than that of \citet{Santos04a} when we
match our results to the data of \citet{Ouchi08}.  We also point out that,
according to what we discussed in \S~\ref{sec:variance}, the simulation box
size used by \citet{Dave06b} (comoving $33\himpc$) was too small to obtain a
reliable LF at the bright-end. Had they used a simulation with a larger box
size and matched their results to the data by \citet{Ouchi08}, we expect that
they would have obtained a value of $\fLya$ similar to ours.  Their galaxy
correlation function at $z=6$ is also lower than ours, probably owing to the
same reason of a small box size.  We note, however, that the galaxy stellar
mass function (in the mass range that they simulated) and the bias in the two
work are consistent with each other.

\begin{table*}[htb]
\begin{center}
\caption{Summary of Two Scenarios}\label{table:summary}
\begin{tabular}{l|cc|cc} \hline \hline\\
Parameter & \multicolumn{2}{c}{Escape Fraction Scenario} & \multicolumn{2}{c}{Stochastic Scenario} \\
 & $z=3$ & $z=6$ & $z=3$ & $z=6$ \\
\hline\\
$\fLya$ $^a$ & 0.10 & 0.15 & $\cdots$ & $\cdots$ \\
$\figm$ $^b$ & 0.82 (1.0) & 0.52 (1.0) & 0.82 (1.0) & 0.52 (1.0) \vspace{0.1cm}\\
$\fion$ $^c$ & 0.06 & 0.20 & $\cdots$ & $\cdots$ \\
$\fdust$ $^d$ & 0.13 (0.11) & 0.36 (0.19) & $\cdots$ & $\cdots$ \\
$\Csto$ $^e$ & $\cdots$ & $\cdots$ & 0.07 (0.06) & 0.2 (0.06) \vspace{0.2cm}\\
$\log (\Mstar/\Msun)$ $^{f}$ & $[10.2, 12.1]$ & $[9.3, 11.0]$ & $[9.2, 11.3]$ & $[8.6, 10.3]$ \vspace{0.1cm}\\
$\langle \Mstar/\Msun \rangle$ $^{g}$ & $2.5\times 10^{10}$ & $1.9\times 10^9$ & $3.8\times 10^9$ & $6.1\times 10^8$ \vspace{0.1cm}\\
$\rho_\star$ fraction $^{h}$ & 0.18 & 0.11 & 0.42 & 0.29 \vspace{0.2cm}\\
$\log (\Mbh/\Msun)$ $^{i}$ & $[7.7, 9.7]$ & $[6.9, 8.7]$ & $[6.7, 8.8]$ & $[6.3, 8.0]$ \vspace{0.1cm}\\
$\langle \Mbh/\Msun \rangle$ $^{j}$ & $1.0\times 10^{8}$ & $9.5\times 10^6$ & $1.5\times 10^7$ & $3.1\times 10^6$ 
\vspace{0.2cm}\\
$\log (Z/\Zsun)$ $^{k}$ & $[-0.7, 0.1]$ & $[-1.3, -0.3]$ & $[-1.0, 0.1]$ & $[-1.4, -0.5]$ \vspace{0.1cm} \\
$\langle Z/\Zsun \rangle$ $^{m}$ & 0.39 & 0.17 & 0.21 & 0.11 \vspace{0.2cm}\\
$r_0$ $^n$ & 5.5 & 4.4 & (3.2) & 3.1 \\
$\gamma$ \footnotemark[$^p$] & 1.67 & 1.68 & (2.30) & 1.49 \\
$b$ $^q$ & 3.5 ($<6.0$) & 5.0 ($<8.9$) & (1.6--4.6) & 4.0 \vspace{0.1cm} \\
\hline \hline\\
\multicolumn{5}{@{}l@{}}{\hbox to 0pt{\parbox{130mm}{\footnotesize
Notes: Summary of various parameter values in the two scenarios discussed in this paper.
\par\noindent
\footnotemark[$a$:] {\em Effective} escape fraction of \Lya\ photons including the effect of IGM attenuation (\S\,\ref{sec:esc}).
\par\noindent
\footnotemark[$b$:] IGM attenuation factor computed by following \citet{Madau95} prescription (\S\,\ref{sec:esc}). The case for $\figm=1.0$ is given in parenthesis.
\par\noindent
\footnotemark[$c$:] Escape fraction of ionizing photons taken from \citet[][Fig.~3]{Inoue06} (\S\,\ref{sec:esc}).
\par\noindent
\footnotemark[$d$:] Extinction of \Lya\ photons by the local dust (\S\,\ref{sec:esc}). The case for $\figm=1.0$ is given in parenthesis.
\par\noindent
\footnotemark[$e$:] Stochasticity (or fractional life time) of LAEs after correcting for the IGM attenuation effect (\S\,\ref{sec:sto}). The case for $\figm=1.0$ is given in parenthesis.
\par\noindent
\footnotemark[$f$:] Stellar mass range of LAEs with $\log \La = 42-44$ (\S\,\ref{sec:meanmass}).
\par\noindent
\footnotemark[$g$:] Mean stellar mass of LAEs with $\log \La = 42-44$ (\S\,\ref{sec:meanmass}).
\par\noindent
\footnotemark[$h$:] Fraction of stellar mass density contained in LAEs with $\log \La = 42-44$ (\S\,\ref{sec:meanmass}).
\par\noindent
\footnotemark[$i$:] Mass range of black holes hosted by LAEs with $\log \La = 42-44$ (\S\,\ref{sec:bh}).
\par\noindent
\footnotemark[$j$:] Mean mass of BHs hosted by LAEs with $\log \La = 42-44$ (\S\,\ref{sec:bh}).
\par\noindent
\footnotemark[$k$:] Stellar metallicity range  LAEs with $\log \La = 42-44$ (\S\,\ref{sec:metal}).
\par\noindent
\footnotemark[$m$:] Mean stellar metallicity of LAEs with $\log \La = 42-44$ (\S\,\ref{sec:metal}).
\par\noindent
\footnotemark[$n$:] Correlation length of LAEs (\S\,\ref{sec:corr}).
\par\noindent
\footnotemark[$p$:] Slope of the auto-correlation function of LAEs (\S\,\ref{sec:corr}).
\par\noindent
\footnotemark[$q$:] Bias parameter of LAEs against dark matter distribution. For the escape fraction scenario, the maximum values at $r\simeq 0.9\,\himpc$ is given in the parenthesis.  The result for $z=3$ stochastic scenario is given in parenthesis, because they are somewhat unreliable owing to small sample size.
}\hss}}
\end{tabular}
\end{center}
\end{table*}

\vspace{0.2cm}

The value of $\fLya=0.02$ assumed by both \citet{Dave06b} and
\citet{Delliou06} is significantly smaller than our value of $\fLya \simeq
0.1$.  It is not easy to understand the source of this difference unless we
compare the SFR function (Fig.~\ref{fig:sfrfunc}) and the effect of dust 
attenuation in each model in detail.  At
least in the case of \citet{Dave06b}, they may have underestimated the value
of $\fLya$ by adjusting the results from a small simulation box size to the
data of \citet{Santos04a}.  
On the other hand, \citet{Kobayashi07} argued for a much larger value 
for the escape fraction, $\fesc \sim 0.8$, by incorporating the effects
of outflows from starburst galaxies. 
As we described earlier, our simple escape fraction model fails to 
reproduce the UV LF of LAEs if their extinction is systematically lower than 
that of LBGs.  It would be interesting to see what the above semi-analytic 
models predict on the joint constraint of the UV LF of LAEs and LBGs at 
$z=3-6$. 

\vspace{0.2cm}

Another source of uncertainty in the present work is the IMF. 
It is possible that the IMF changes as a function of redshift, 
metallicity, and environment \citep[e.g.,][]{Larson98}.  
Recently there have been several suggestions 
that some of the observational data can be better explained if the IMF 
becomes ``top-heavy'' or ``bottom-light'' towards \highz\
\citep[e.g.,][]{Baugh05, Chary07, Dokkum07a}.  
It would also mitigate the apparent discrepancy between the data on extinction 
corrected SFR density and stellar mass density 
\citep[e.g.,][]{Nag04d, Nag06c, Fardal07, Dave08, Wilkins08}. 
If indeed the IMF is top-heavy at \highz, then the conversion factor
in Eq.~(\ref{eq:sfr}) could increase by a factor of $\sim 2$ \citep{Schaerer03}, 
and our intrinsic \Lya\ LF would be brighter by the same factor. 
In this case, the values of $\fLya$ and $\Csto$ should be decreased 
by similar factors. 
The top-heavy IMF would also disrupt the agreement of the UV LF of LBGs
at $z=3$ between simulations and the observed data, and exacerbate
the discrepancy between the simulation and the observed data at $z=6$. 
As far as the \Lya\ \& UV LFs of LAEs are concerned, we do not see any 
strong reasons to invoke a top-heavy IMF based on our comparisons,
although we cannot rule out the possibility of top-heavy IMF 
given various uncertainties discussed in this paper. 

\bigskip

This work is supported in part by the NSF grant AST-0807491, 
National Aeronautics and Space Administration under Grant/Cooperative 
Agreement No. NNX08AE57A issued by the Nevada NASA EPSCoR program, and 
the President's Infrastructure Award from UNLV. 
KN is grateful for the hospitality of IPMU, University of Tokyo, and 
the Aspen Center for Physics, where part of this work was done. 
The simulations and analyses for this paper were 
performed at the Institute of Theory and Computation at 
Harvard-Smithsonian Center for Astrophysics and UNLV Cosmology
Computing Cluster.

\end{document}